\documentclass[twocolumn,floats,floatfix,nofootinbib,aps,pra]{revtex4-2}   
\usepackage{amsfonts,amssymb,amsmath,amsbsy}
\usepackage{physics}
\usepackage{color,calc}
\usepackage{bm}
\usepackage{array}
\usepackage{booktabs}
\usepackage{graphicx}
\usepackage{graphics}
\usepackage{eepic,epsfig}
\usepackage[breaklinks,hidelinks,colorlinks=true,linkcolor=blue,citecolor=blue,filecolor=blue,urlcolor=blue]{hyperref}

\usepackage[dvipsnames]{xcolor}
\definecolor{myblue}{RGB}{0, 0, 255}
\usepackage[toc,page]{appendix}

\def\be{ \begin{equation} }
\def\ee{ \end{equation} }
\def\bea{ \begin{eqnarray} }
\def\eea{ \end{eqnarray} }
\def\bse{ \begin{subequations} }
\def\ese{ \end{subequations} }
\def\ba{ \begin{array} }
\def\ea{ \end{array} }
\def\bt{ \begin{tabular} }
\def\et{ \end{tabular} }

\def\i{\,\text{i}}
\def\e{\,\text{e}}
\def\i{i}
\def\e{e}

\def\to{\rightarrow}

\def\d{\text{d}}

\def\U{\mathbf{U}}

\def\A{\mathcal{A}}

\def\i{{\rm{i}}}
\def\f{{\rm{f}}}
\def\phase{\phi}

\long\def\/*#1*/{}

\begin{document}

\title{Narrowband and passband composite rotational quantum gates}

\author{Hayk L. Gevorgyan\textsuperscript{\hyperref[1]{1},\hyperref[2]{2},\hyperref[3]{3}}}
\affiliation{\phantomsection\label{1}{\textsuperscript{1}Center for Quantum Technologies, Faculty of Physics, St. Kliment Ohridski University of Sofia, 5 James Bourchier Blvd., 1164 Sofia, Bulgaria}\\
\phantomsection\label{2}{\textsuperscript{2}Quantum Technologies Division, Alikhanyan National Laboratory (Yerevan Physics Institute), 2 Alikhanyan Brothers St., 0036 Yerevan, Armenia}\\
\phantomsection\label{3}{\textsuperscript{3}Experimental Physics Division, Alikhanyan National Laboratory (Yerevan Physics Institute), 2 Alikhanyan Brothers St., 0036 Yerevan, Armenia}}
\email{hayk.gevorgyan@aanl.am}

\date{\today }

\begin{abstract}

High-precision, robust quantum gates are essential components in quantum computation and information processing. In this study, we present an alternative perspective, exploring the potential applicability of quantum gates that exhibit heightened sensitivity to errors. We investigate such sensitive quantum gates, which, beyond their established use in in vivo NMR spectroscopy, quantum sensing, and polarization optics, may offer significant utility in precision quantum metrology and error characterization. Utilizing the composite pulses technique, we derive three fundamental quantum gates with narrowband and passband characteristics --- the X (NOT) gate, the Hadamard gate, and gates enabling arbitrary rotations. To systematically design these composite pulse sequences, we introduce the SU(2), modified-SU(2), and regularization random search methodologies. These approaches, many of which are novel, demonstrate superior performance compared to established sequences in the literature, including NB1, SK1, and PB1.

\end{abstract}

\maketitle



\section{Introduction\label{Sec:intro-ch4}}
Quantum rotation gates \cite{nielsen2000, chen2006}, are the key elements in experimental quantum computing. Interestingly, a Rabi rotation gate, being SU(2), is in the heart of various quantum computing devices, especially AMO (atomic, molecular, optical), while suggested theoretical quantum gates are U(2). Furthermore, a quantum circuit designed by multiple quantum gates represents a composite rotation on a Bloch sphere. X (NOT) and Hadamard rotation gates are two special cases of a single $\theta$ pulse or rotation, when $\theta = \pi$ and $\theta = \pi/2$, respectively. On contrary, phase gates require at least two rotations. A method that reveals the benefits of composite rotation gates is composite pulses (CPs).          

Although CPs, first, have been used in polarization optics (PO) \cite{pol, pol1}, the name, classification and development of the technique belongs to the area of nuclear magnetic resonance (NMR) \cite{nmr, nmr1, nmr2, wimperis1990, wimperis1994, variable-constant-1, variable-constant-2}. Being efficient and versatile control technique, CPs may easily adapt to various requirements. This feature manifests in the wide range of applications in both quantum and classical physics --- qubit control in trapped ions \cite{trapped-ions-teleportation, trapped-ions, trapped-ions1, trapped-ions2, trapped-ions3, trapped-ions4, trapped-ions5}, neutral atoms \cite{neutral-atoms}, doped solids \cite{doped-solids, doped-solids1, doped-solids2}, NV centers in diamond \cite{nv-centers, nv-centers-entanglement}, and quantum dots \cite{quantum-dots, quantum-dots1, quantum-dots2, quantum-dots3}, high-accuracy optical clocks \cite{clocks}, cold-atom interferometry \cite{cold-atom, cold-atom1, cold-atom2}, optically dense atomic ensembles \cite{atomic-ensembles}, magnetometry \cite{magnetometry}, optomechanics \cite{optomech}, Josephson junctions \cite{josephson}, magnetic resonance imaging (MRI) \cite{mri}, NMR quantum computation (NMR QC) \cite{nmr-qc, nmr-qc-1}, entanglement generation \cite{nv-centers-entanglement}, teleportation \cite{teleportation, teleportation1, trapped-ions-teleportation}, molecular spectroscopy \cite{mol-spectr} etc. The possibility of applying a deep neural network for design of CPs is distinguished by its modernity \cite{neural}.   

A composite pulse sequence is a finite train of pulses with specified pulse areas ($\theta$-s) and relative phases ($\phi$-s), and in a specific order. Since the initial target in NMR was a composite $\pi$ pulse with the structure known beforehand (requires a sequence of $\pi$ pulses), composite phases were in the foreground. Considering various engineering perspectives, it is common to divide CPs into the three main branches --- broadband (BB), narrowband (NB) and passband (PB) classes, given by Wimperis \cite{wimperis1990, wimperis1994}. 



From the point of view of mathematics, the CPs can be represented as composite rotations on the Bloch-Poincaré sphere. This leads to the second kind of classification of CPs --- \emph{variable} (class B) and \emph{constant} (phase-distortionless, fully-compensating or class A) \emph{rotations}, given by Levitt \cite{variable-constant-1, variable-constant-2, variable-constant-3}. 

Constant rotation CPs are independent of the initial state and not permit distortions of the phase of the overall propagator in the rotation axis over a wide error band, if not over the entire error range. Combining in one word, they are ``universal'' over the entire Bloch sphere, which, for instance, makes them applicable to quantum computation \cite{gevorgyan2021, gevorgyan2024}. In NMR and magnetic resonance imaging (MRI), constant rotations are often used in advanced, phase-sensitive (require phase cycling) two-dimensional NMR experiments, like COSY \cite{COSY} and TOCSY, providing a powerful tool for the determination of the chemical structure of molecules.

BB CPs act as \texttt{TARGET} operator in a broad range of errors around $0$ value (flat-top fidelity), and as \texttt{IDENTITY} operator only at $\pm 1$, which expands the fidelity profile, when NB CPs act as \texttt{TARGET} only at $0$, and as \texttt{IDENTITY} at the ranges left to $1$ and right to $-1$ (flat-bottom fidelity), which squeezes the fidelity profile. PB CPs merge these two properties, i.e., they are expanded at the center and squeezed at the edges. Longer sequences can enhance the property: designing broader or/and narrower fidelity profiles. 

Also it is possible to enhance the property more, using (NB2, BB2, PB2)-like CPs, which we call \emph{ultra-sequences} (ultra-BB, ultra-NB, ultra-PB). These CPs improve with a loss in precision that now fluctuates within a certain range and is no longer flat. Ultra-sequences are applicable in areas where ultrahigh-precision is optional \cite{gevorgyan2022, gevorgyan2024arxiv, gevorgyanthesis2023}.

Constant rotations \cite{gevorgyan2021, gevorgyan2024} are obviously more demanding than variable ones \cite{Torosov2019, torosov2020} and require longer sequences for the same order of compensation, as they require propagator-optimization (or so-called full-compensation)\footnote{Note, that the naming ``constant rotation'' is true for broadband composite pulses, since for the narrowband composite pulses the phase $\phi(\epsilon)$ is now more sensitive rather stable (robust). Therefore, we use the terms propagator (gate) optimization and full optimization.}. Previously, this was theoretically done using the theory of the average Hamiltonian, the Magnus expansion, or the theory of quaternions. Using theoretical way, Wimperis found BB1, NB1 and PB1 rotations \cite{wimperis1994}, which have second order of optimization. In the following papers narrowband or/and passband composite pulses are considered \cite{Torosov2023, Zhang2024}. The authors of the article \cite{Mallweger2024} offer and demonstrate a new method for measuring ion movement in the quantum regime, i.e., close to absolute zero. To the best of our knowledge, the first consideration to theoretically design the narrowband composite sequences of arbitrary length is done in \cite{vitanov2011, Ivanov2011, Torosov2011, Torosov2011g}.

We propose previously unseen classes of composite pulses, namely, narrowband and passband composite pulses with propagator-optimization, i.e., with full compensation of both major and minor diagonal matrix elements. This kind of compensation makes both overall pulse area and overall phase of the rotation gate sensitive\footnote{For narrowband case.} (and robust\footnote{For passband case.}), making fully-optimized narrowband (fully sensitive) rotation gate. We propose SU(2) random search method and it's alternatives to numerically generate these rotations in a structured way: searching for all candidates in accordance with the compensation order.  

This paper is organized as follows. In Sec.~\ref{Sec:derivation} we explain the derivation methods. Composite X gates are presented in Sec.~\ref{Sec:x}, while composite Hadamard gates in Sec.~\ref{Sec:h}. Sec.~\ref{Sec:gen} is devoted to the general rotation gates. The last-mentioned three sections are divided into two subsections, presenting narrowband and passband rotations. Finally, Sec.~\ref{Sec:concl} presents the conclusions.










\section{Derivation \label{Sec:derivation}}

We derive rotation gates similar to the previous work \cite{gevorgyan2021}. We are dealing with the following propagator with SU(2) symmetry
\be\label{SU(2)}
\U_0 = \left[ \begin{array}{cc} a & b \\ -b^{\ast} & a^{\ast} \end{array}\right],
\ee
where $a$ and $b$ are the complex-valued Cayley-Klein parameters satisfying $|a|^2+|b|^2=1$, where $a=\cos(\A/2) $ and $b=-\i\sin(\A/2)$. $\A$ represents 
the temporal pulse area $\A=\int_{t_\i}^{t_\f}\Omega(t)\d t$ in quantum optics, the pulse width or amplitude $\theta$ in NMR, and the phase shift $\varphi = 2 \pi L (n_f - n_s)/\lambda$ \cite{rangelov2015} in polarization optics. Without loss of generality of the problem, we will use the terminology of quantum computing. 

\begin{figure}[t]
\bt{r}
\centerline{\includegraphics[width=0.8\columnwidth]{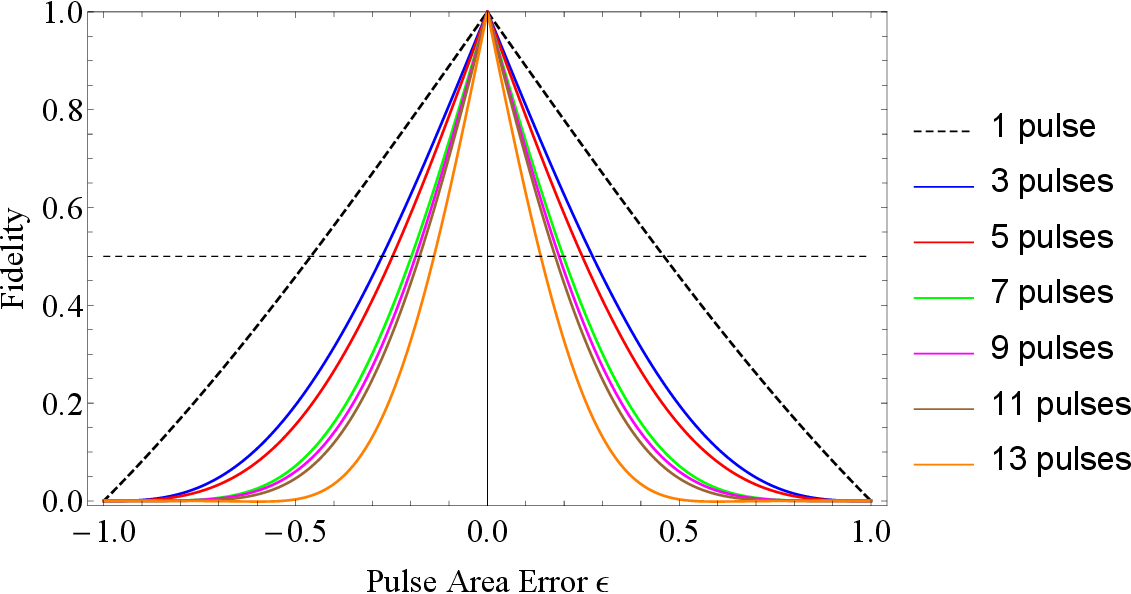}} \\ \\
\centerline{\includegraphics[width=0.8\columnwidth]{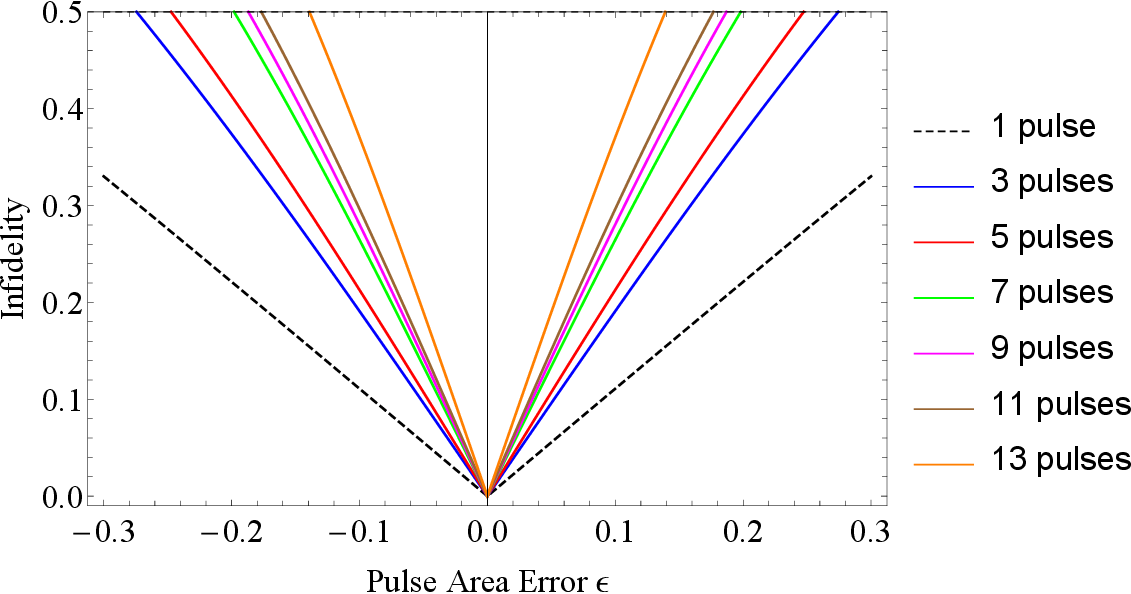}}
\et
\caption{
Frobenius distance fidelity (top) and infidelity (bottom) of composite narrowband X gates produced by the antisymmetric composite sequences A$N$s (3,7,11 pulses) and W$N$s (5,9,13 pulses) from the Table~\ref{Table:NB-X-AN}.
}
\label{fig:NB-X-AN}
\end{figure}

A phase shift $\phase$ imposed on the driving field, $\Omega(t)\to\Omega(t)\e^{\i\phase}$, is imprinted onto the propagator \eqref{SU(2)} as
\be\label{U phase}
\U_\phase = \left[ \begin{array}{cc} a & b \e^{\i\phase} \\ -b^{\ast}\e^{-\i\phase} & a^{\ast} \end{array}\right].
\ee
An explicit formula of \eqref{U phase} is a (Rabi) rotation gate
\be\label{U phase 1}
\U_\phase = \left[ \begin{array}{cc} \cos{(\A/2)} & - i \sin{(\A/2)} \e^{\i\phase} \\ - i \sin{(\A/2)} \e^{-\i\phase} & \cos{(\A/2)} \end{array}\right],
\ee
which can be represented by Pauli matrices
\be\label{U phase 2}
\U_\phase = e^{ - i \A \left(\sigma_x \cos{\phase} - \sigma_y \sin{\phase} \right)}.
\ee

Composite pulses require chronological action of evolutions of the \eqref{U phase 1} type. A train of $N$ pulses, each with area $\A_k$ and phase $\phase_k$ (applied from left to right),
\be
(\A_1)_{\phi_1} (\A_2)_{\phi_2} (\A_3)_{\phi_3} \cdots (\A_N)_{\phi_{N}},
\ee
produces the propagator (acting, as usual, from right to left)
\be\label{U^N}
\boldsymbol{\mathcal{U}} = \U_{\phase_{N}}(\A_N) \cdots \U_{\phase_{3}}(\A_3) \U_{\phase_{2}}(\A_2) \U_{\phase_{1}}(\A_1).
\ee

Under the assumption of a single systematic error of pulse area $\epsilon$ (all the pulse areas are errant $\A_k \rightarrow \A_k (1+\epsilon)$), we can expand the composite propagator \eqref{U^N} in a Taylor series versus $\epsilon$.
Because of the SU(2) symmetry of the overall propagator, it suffices to expand only two of its elements, say $\mathcal{U}_{11}(\epsilon)$ and $\mathcal{U}_{12}(\epsilon)$. Since our \texttt{TARGET} operator is the so-called $\theta$ pulse (with phase $\phi = \pi/2$) or $\theta$ rotation gate, i.e.
\be\label{target}
\texttt{TARGET} \overset{\Delta}{=} \vb{R}(\theta) = e^{i \theta \sigma_y /2},
\ee
we set their zero-error values to the target values,
\be\label{eq-0}
\mathcal{U}_{11}(0) = \cos(\theta/2),\quad \mathcal{U}_{12}(0) = \sin(\theta/2).
\ee

Rotation gates require propagator-optimization, hence, we will optimize both major $\mathcal{U}_{11}(\epsilon)$ and minor $\mathcal{U}_{12}(\epsilon)$ diagonal elements.

\subsection{Narrowband composite pulses}
\subsubsection{SU(2) approach}
Here, we set as many of their derivatives with respect to $\epsilon$ at $\pm 1$, in the increasing order, as possible,
\be\label{eq-s}
\mathcal{U}^{(m)}_{11}(\pm) = 0,\quad\mathcal{U}^{(m)}_{12}(\pm) = 0, \quad (m=1,2,\ldots, n_s),
\ee
where $ \mathcal{U}^{(m)}_{jl} = \partial_\epsilon^m  \mathcal{U}_{jl}$ denotes the $m$th derivative of $\mathcal{U}_{jl}$ with respect to $\epsilon$.
The largest derivative order $n_s$ satisfying Eqs.~\eqref{eq-s} gives the order of sensitivity $O(\epsilon^{n_s})$.

Derivation of the NB CPs requires the solution of Eqs.~\eqref{eq-0} and \eqref{eq-s}. We do this numerically by using standard routines in \textsc{Mathematica}: we minimize the following loss or error function of optimization 
\be\label{method-1}
\begin{split}
\mathcal{E} &= \mathcal{E}_0 + \\
&+ \sum\limits_{k=1}^{n_s} \Bigr[\lvert\mathcal{U}^{(k)}_{11}(-) \rvert^2 + \lvert\mathcal{U}^{(k)}_{11}(+) \rvert^2 + \lvert\mathcal{U}^{(k)}_{12}(-)\rvert^2 + \lvert\mathcal{U}^{(k)}_{12}(+)\rvert^2 \Bigr], 
\end{split}
\ee
where the initial condition (targeted gate) is captured by $\mathcal{E}_0 = \lvert \mathcal{U}_{11}(0) - \cos{\theta/2} \rvert^2 + \lvert \mathcal{U}_{12}(0) - \sin{\theta/2} \rvert^2$, and $n_s$ is the narrowness or sensitivity order. This minimization method is similar to the least squares, since the sum of the absolute squares of the components of error function are taken. Random search of the minimum of this form provides fast results. For example, the well-known NB1 CP can be derived by using SU(2) approach.    

\begin{figure}[t]
\bt{r}
\centerline{\includegraphics[width=0.8\columnwidth]{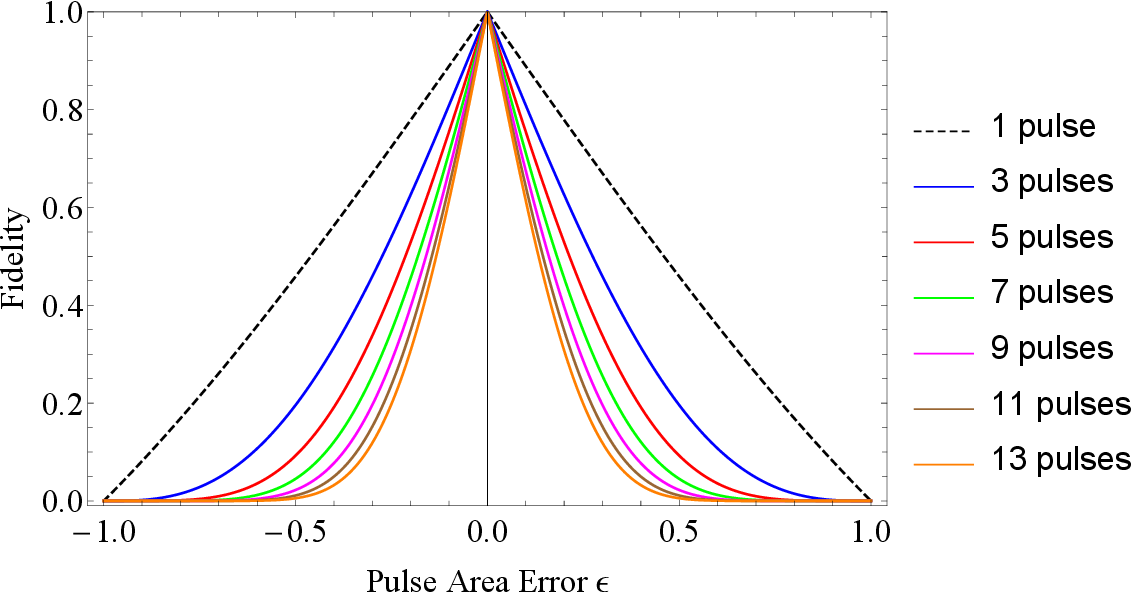}} \\ \\
\centerline{\includegraphics[width=0.8\columnwidth]{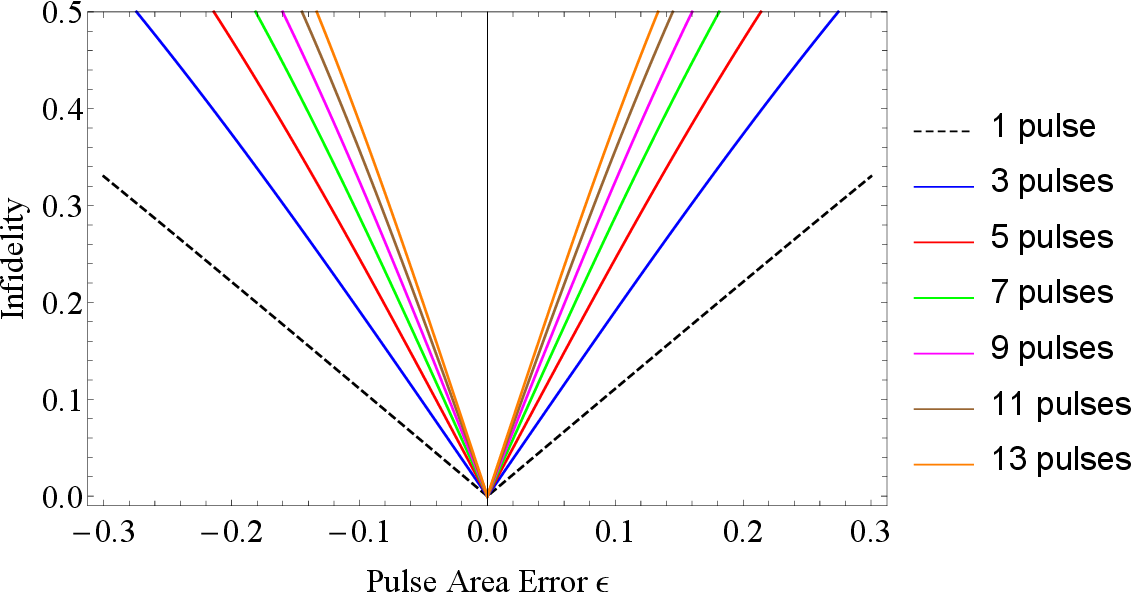}}
\et
\caption{
Frobenius distance fidelity (top) and infidelity (bottom) of composite X gates produced by the antisymmetric composite sequences A$N$-m designed by the regularization method from the Table~\ref{Table:NB-X-ANm}.
}
\label{fig:NB-X-ANm}
\end{figure}

\subsubsection{Modified-SU(2) approach}
We have noticed that sometimes it is better to use the modified version of SU(2) approach. Major and minor diagonal elements of SU(2) matrix are related $\lvert\mathcal{U}_{11}(\epsilon)\rvert^2 + \lvert\mathcal{U}_{21}(\epsilon)\rvert^2 = 1$, being Cayley-Klein parameters. Due to this dependence, optimization of one will directly narrower the other one. To ensure the sensitivity of the phase of the narrowband rotation, we optimize the minor diagonal element. So, the loss function is the following 
\be\label{method-2}
\begin{split}
\mathcal{E} = \mathcal{E}_0 + \sum\limits_{k=1}^{2 n_s} \Bigr[ \lvert\mathcal{U}^{(k)}_{12}(-)\rvert^2 + \lvert\mathcal{U}^{(k)}_{12}(+)\rvert^2 \Bigr].
\end{split}
\ee
Modified SU(2) approach works for X gate or $\pi$ rotations, and gives better results than by using SU(2). The number of derivatives optimized by both methods are equal, but by this method the minor element (of the actual gate matrix) $\mathcal{U}_{21}(\epsilon)$ is optimized by the order of $2 n_s$, two times the sensitivity order. The major element vs error $\mathcal{U}_{11}(\epsilon)$ dependence is already sharp-narrow and symmetric, but minor element vs error $\mathcal{U}_{21}(\epsilon)$ is bell-shaped.

Unfortunately, this method does not work for Hadamard gate or $\pi/2$ rotations. The reason is an asymmetry in both major and minor element dependences $\mathcal{U}_{11}(\epsilon)$ and $\mathcal{U}_{21}(\epsilon)$, which can not be modified to be symmetric. One side is easier to optimize than the other side. The optimization of $\mathcal{U}_{11}(\epsilon)$ at negative side $\epsilon=-1$ is easier than at positive side $\epsilon=1$, and vice versa for $\mathcal{U}_{12}(\epsilon)$. Maybe there is a method which can do asymmetric optimization which will give better results for Hadamard gate in fidelity representation, but we limit ourselves to the SU(2) approach, since for rotation gate both sides are important.

\subsection{Passband composite pulses}
\subsubsection{SU(2) approach}
As already mentioned, PB composite pulses have the properties of both BB and NB CPs. In addition to the narrowband property \eqref{eq-s}, we add broadband property 
\bse\label{eq-p}
\begin{gather}
\mathcal{U}^{(k)}_{11}(0) = 0,\quad\mathcal{U}^{(k)}_{12}(0) = 0, \quad (k=1,2,\ldots, n_r), \label{eq-p-r} \\
\mathcal{U}^{(m)}_{11}(\pm) = 0,\quad\mathcal{U}^{(m)}_{12}(\pm) = 0, \quad (m=1,2,\ldots, n_s). \label{eq-p-s}
\end{gather}
\ese
Now, in addition to sensitivity order $n_s$ in Eq.~\eqref{eq-p-s}, we also have $n_r$ which is the largest derivative order satisfying Eq.~\eqref{eq-p-r} and gives the order of robustness $O(\epsilon^{n_r})$. Pulse sequence with any combination of $n_s$ and $n_r$ both equal or greater than one is passband. Therefore, we examine two types of passband CPs, namely
\begin{itemize}
 \item \emph{pari passu}\footnote{In Latin, \emph{pari passu} means ``with equal step'' or ``on equal footing.'' By this naming, we mean the equality of the orders of properties (robustness and sensitivity).} passband composite pulses, for which robustness and sensitivity orders are equal and define the passband order $n_p = n_r = n_s$,
 \item \emph{diversis passuum}\footnote{In Latin, \emph{diversis passuum} translates to ``of different steps'' or ``of diverse paces.'' By this naming, we mean the difference between the orders of properties (robustness and sensitivity).} passband composite pulses, for which one of the above properties is superior to the other $n_r \neq n_s$.
\end{itemize}

Derivation of the PB CPs requires the solution of Eqs.~\eqref{eq-0}, and \eqref{eq-p}. We do this numerically by using standard routines in \textsc{Mathematica}: we minimize the following loss function of optimization
\be\label{method-pb}
\begin{split}
\mathcal{E} & = \mathcal{E}_0 + \sum\limits_{k=1}^{n_r} \Bigr[\lvert\mathcal{U}^{(k)}_{11}(0) \rvert^2 + \lvert\mathcal{U}^{(k)}_{12}(0) \rvert^2\Bigr] + \\
& + \sum\limits_{k=1}^{n_s} \Bigr[\lvert\mathcal{U}^{(k)}_{11}(-) \rvert^2 + \lvert\mathcal{U}^{(k)}_{11}(+) \rvert^2 + \lvert\mathcal{U}^{(k)}_{12}(-)\rvert^2 + \lvert\mathcal{U}^{(k)}_{12}(+)\rvert^2 \Bigr]. 
\end{split}
\ee
For example, the well-known SK1 ($n_p = 1$) (Solovay-Kitaev method \cite{SK1}) and PB1 ($n_p = 2$) (Wimperis \cite{wimperis1994}) and composite pulses can be derived by using SU(2) approach, which, of course, are pari passu. But for $n_p \geq 3$, this straightforward cancellation of required derivatives in both major and minor elements result in the alternating or wiggled composite pulses. In our opinion, the composite pulse can't take on such a precise optimization --- due to this inflexible method, the composite pulse tends to be more square than possible, and these wiggles occur. For this reason, we use more flexible method of the propagator-optimization, which we call the method of regularization. Despite this, this method is useful for obtaining diversis passuum composite pulses (not the longer ones since the wiggles occur in the same sense).

\begin{figure}[t]
\bt{r}
\centerline{\includegraphics[width=1\columnwidth]{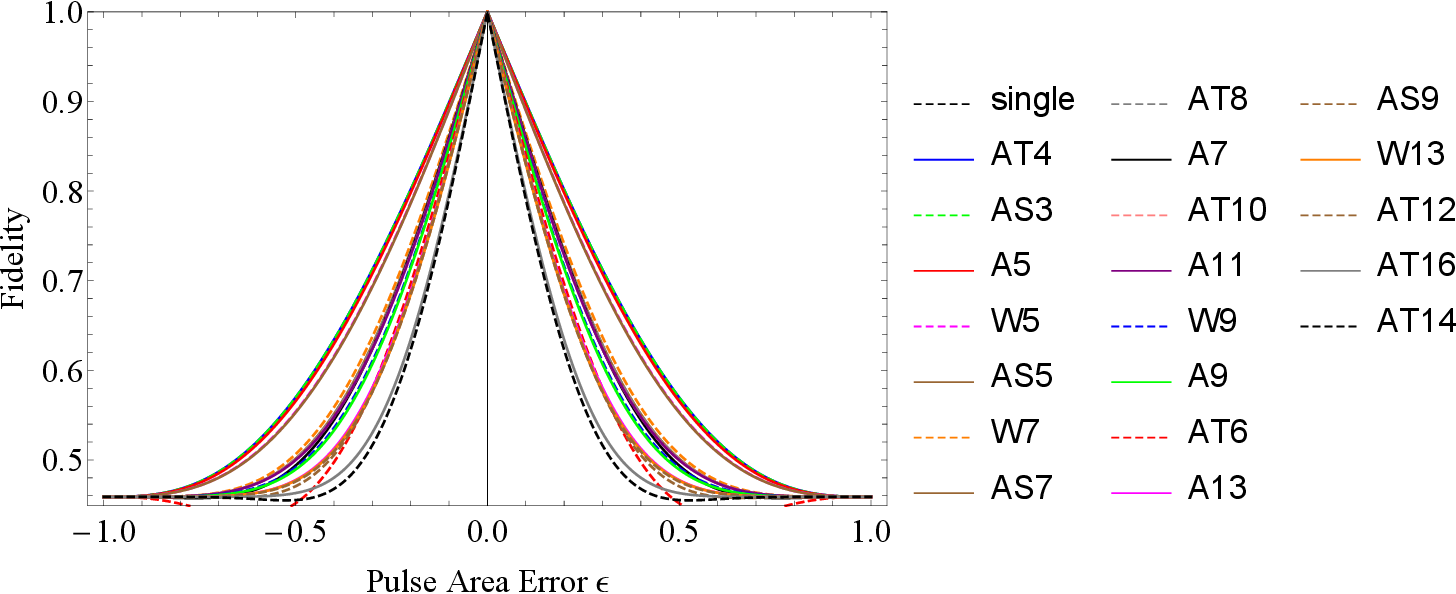}} \\ \\
\centerline{\includegraphics[width=1\columnwidth]{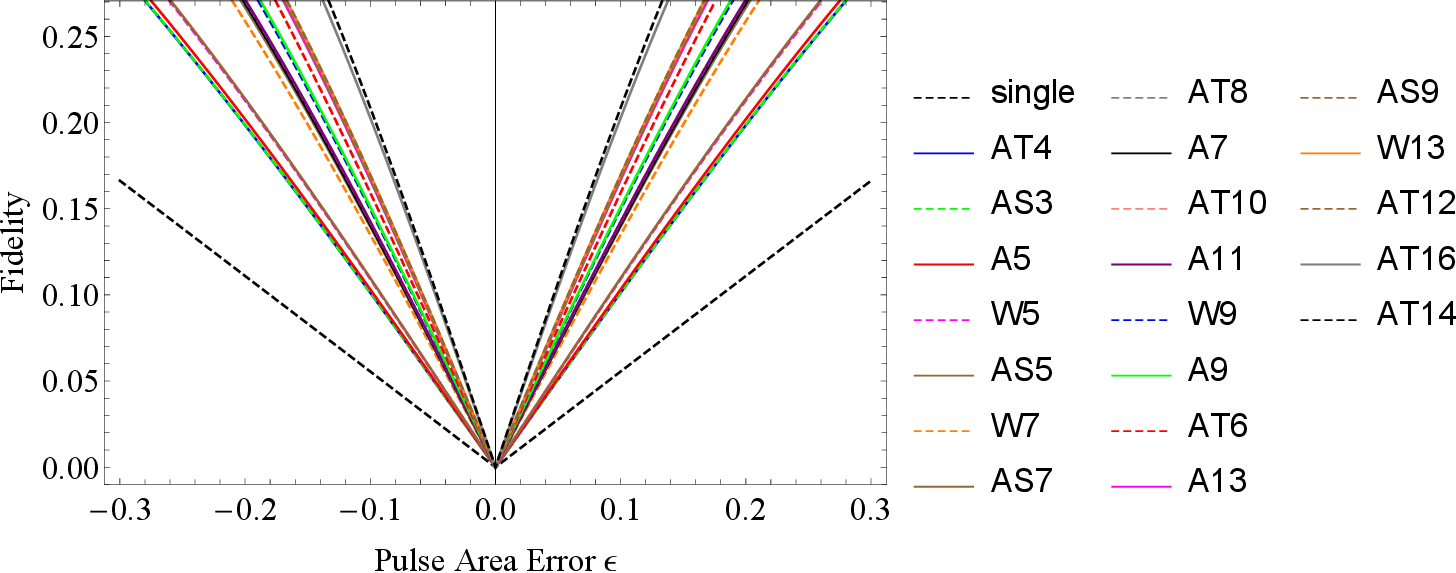}}
\et
\caption{
Frobenius distance fidelity (top) and infidelity (bottom) of composite narrowband Hadamard gates produced by the four families of composite sequences from the Table~\ref{Table:NB-H}.
}
\label{fig:NB-H}
\end{figure}

\subsubsection{Regularization approach}
Results obtained by the SU(2) method of derivation, besides SK1 and PB1, have wiggles on the edges, arising negative fidelity. The optimization method, alternative to SU(2), is more flexible and gives better results is a regularization method 
\be\label{method-reg}
\begin{split}
\mathcal{E} & = \mathcal{E}_0 + \sum_{k=1}^{2 n_p} \Bigr[ \lvert\mathcal{F}_{\text{T}}^{(k)} (0)\rvert^2 + \lvert\mathcal{F}_{\text{T}}^{(k)} (-)\rvert^2 + \lvert\mathcal{F}_{\text{T}}^{(k)} (+)\rvert^2 \Bigr] + \\
& + \lambda \Bigr[ \lvert\mathcal{U}^{\prime}_{11}(0) \rvert^2 + \lvert\mathcal{U}^{\prime}_{12}(0) \rvert^2 + \lvert\mathcal{U}^{\prime}_{11}(+) \rvert^2 + \\
& + \lvert\mathcal{U}^{\prime}_{11}(-) \rvert^2 + \lvert\mathcal{U}^{\prime}_{12}(+) + \lvert\mathcal{U}^{\prime}_{12}(-) \rvert^2 \Bigr],
\end{split}
\ee
where $2 n_p$ orders of narrowness/broadness of trace fidelity of SU(2) matrix are optimized, which is equivalent to the optimization of SU(2) matrix elements by the order of $n_p$ (two times lower). A regularizer $\lambda \neq 0$ constrains the result to be fully-optimized and without unnecessary wiggles. In our optimization, it is taken $\lambda = 1$. As you may notice, we constrain ourselves to deriving the pari passu CPs using the regularization method, although it can also be used to derive the diversis passuum CPs. The aim of our work is to show the diversity of the CPs and how to derive them.

\subsection{Performance measures}
As in our previous works \cite{gevorgyan2021, gevorgyan2024}, here we also use the Frobenius distance fidelity,
\be\label{Frobenius}
\mathcal{F} = 1 - \| \boldsymbol{\mathcal{U}} (\epsilon) - \mathbf{R}(\theta) \|
 = 1 - \sqrt{ \tfrac14 \sum\nolimits_{j,k=1}^2 \left|\mathcal{U}_{jk} - R_{jk} \right|^2 },
\ee
as the measure of performance of rotation gates. Alternatively, since this is a common practice in the NMR quantum computing community, the trace fidelity can be used,
\be\label{trace fidelity}
\mathcal{F}_{\text{T}} = \tfrac12 \text{Tr}\, [ \boldsymbol{\mathcal{U}} (\epsilon) \mathbf{R}(\theta)^\dagger ].
\ee

Since we consider full optimization, as for constant rotations, the fidelities can not be negative\footnote{Note, that for both fidelities we didn't take absolute values, therefore, we can have negativity in general. Certainly, one must take the absolute value when to show the precision measure, but this is more of convention.}, while it can be the case for composite pulses alternating at the bottom, i.e. like NB2 and PB2. This problem of negative fidelities could be solved by taking absolute value as was done for the trace fidelity, but we suggest not to take since it has meaning which can be neglected in the opposite case. Negative fidelity means that actual and target matrices are so far from each other, that the infidelity or the norm $\| \vb*{\mathcal{U}} (\epsilon) - \vb*{R}(\theta) \| = \sqrt{ \tfrac14 \sum\nolimits_{j,k=1}^2 \lvert\mathcal{U}_{jk} - R_{jk} \rvert^2}$, the square root of the sum of the squares of absolute values of closeness of matrix elements, is greater than 1. In the worst case, when actual and target matrices have opposite signs, the possible minimum values of fidelities are $-1$ and $1-\sqrt{2}$ for trace and distance fidelities, respectively. Opposite sign represents inessential global phase for quantum gates, and taking absolute value of the fidelity is associated with this consideration. Anyway, this is not our case, since we do not have alternations or wiggles of the fidelity and the minimum of fidelity is at it's boundaries $\epsilon = \pm 1$. As we notice, maximum or top fidelity is equal $100\%$ and represents pure correspondence of actual and target gates $\left[\mathcal{F} (\theta) \right]_{max} = \left[\mathcal{F} (\theta) \right]_{\epsilon = 0} = 1$. Despite this, the minimum or bottom fidelity depends on the $\theta$ parameter  
\bse\label{30}
\begin{align}
& \left[\mathcal{F}_T (\theta) \right]_{min} = \left[\mathcal{F}_T (\theta) \right]_{\epsilon = \pm 1}  = \cos{\frac{\theta}{2}}, \\
& \left[\mathcal{F} (\theta) \right]_{min} = \left[\mathcal{F} (\theta) \right]_{\epsilon = \pm 1}  = 1 - \sqrt{1-\cos{\frac{\theta}{2}}}.
\end{align}
\ese
For the X gate both fidelity measures are zero at the bottom $\left[\mathcal{F} (\pi) \right]_{min} = \left[\mathcal{F}_T (\pi) \right]_{min} = 0$, when for the Hadamard gate $\left[\mathcal{F} (\pi/2) \right]_{min} = 1 - \sqrt{1 - \frac{1}{\sqrt{2}}}$, $\left[\mathcal{F}_T (\pi/2) \right]_{min} = \frac{1}{\sqrt{2}} $ they are greater than zero and differ from each other. Moving from $\pi + \pi k$ rotations to the $\pi/2 + \pi k$ ($\forall k \in \mathbb{Z}$), the bottom fidelity increases. Hence, the performances of different rotation gates can not be compared perfectly. This is a drawback of the fidelity measures. Especially, for sensitivity (narrowness of fidelity) measures, the presence of a bottom fidelity can not be neglected. We calculate the full width at half-maximum (FWHM) for narrowband composite rotation gates as $\mathcal{F}_{HM} = \frac{\left[\mathcal{F}\right]_{max} + \left[\mathcal{F} (\theta) \right]_{min}}{2} = \frac{1 + \left[\mathcal{F} (\theta) \right]_{min}}{2}$, where $\left[\mathcal{F} (\theta) \right]_{min}$ is a bottom fidelity. Likewise, UL-fidelity (ultralow) is computed by adding the value of $10^{-4}$ to the bottom fidelity $\left[\mathcal{F} (\theta) \right]_{min}$, when UH-fidelity (ultrahigh) were computed by substituting this value from the top fidelity 1. 

\begin{figure}[t]
\bt{r}
\centerline{\includegraphics[width=0.8\columnwidth]{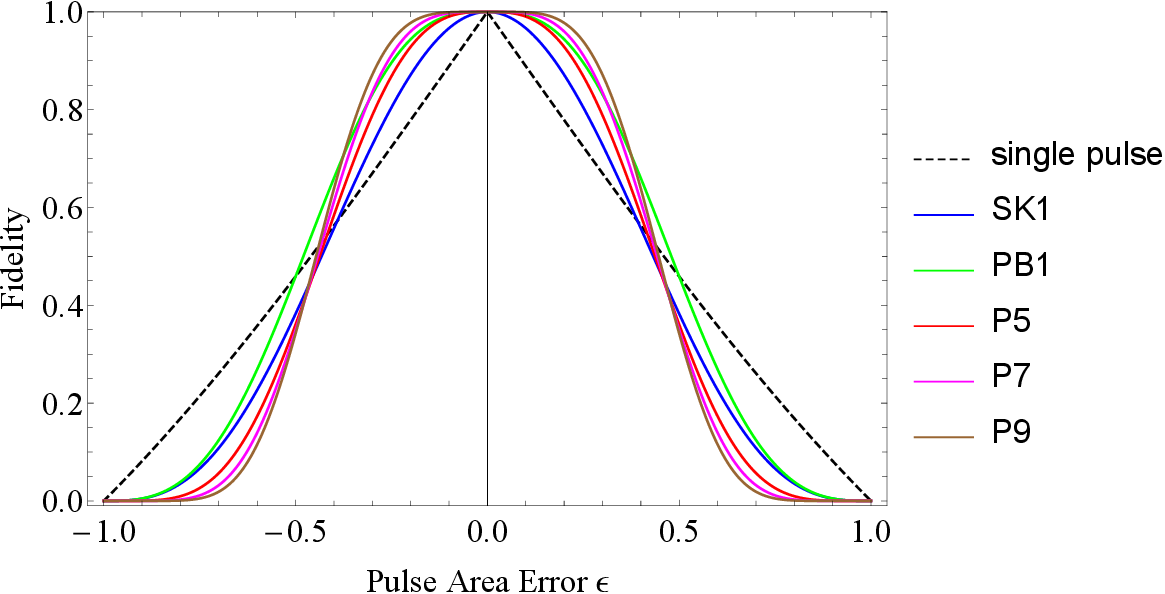}} \\ \\
\centerline{\includegraphics[width=0.8\columnwidth]{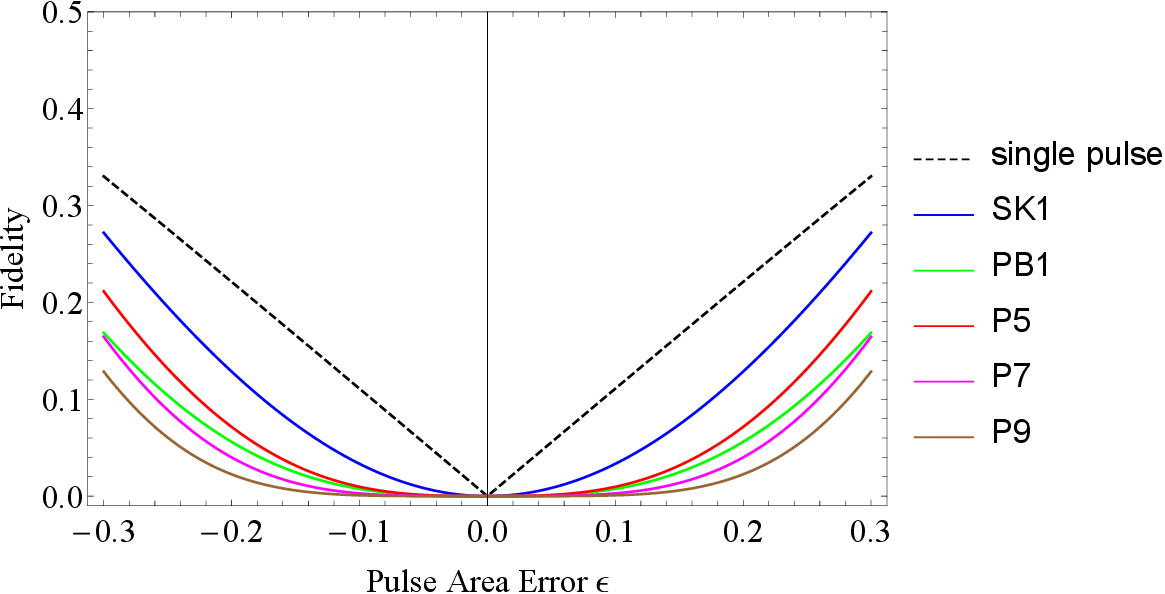}}
\et
\caption{
Frobenius distance fidelity (top) and infidelity (bottom) of composite passband X gates produced by P$N$ (pari passu) sequences from the Table~\ref{Table:PB-X-PN}.
}
\label{fig:PB-X-PN}
\end{figure}

We propose to use the measure $\Delta(\alpha_0) = \lvert\epsilon\left(\mathcal{F} = \alpha_0\right)\rvert -\lvert\epsilon\left(\mathcal{F} = 1-\alpha_0\right)\rvert$ of the rectangularity of passband CPs. In our case, for rotation gates, we choose  $\alpha_0$ equal to $10^{-4}$, which corresponds to the quantum computation benchmark, and rectangularity measure $\Delta \overset{\Delta}{=} \Delta(10^{-4})$ is the difference between absolute errors at UL-(ultralow) and UH-fidelities (ultrahigh). Since the slope coefficient (is approximated by a straight line $\tan{\beta_0} \simeq \frac{\Delta{\mathcal{F}}}{\Delta(\alpha_0)} = \frac{1-2\alpha_0}{\Delta(\alpha_0)} $) is inversely proportional to $\Delta$, hence, smaller $\Delta$, higher the rectangularity of the fidelity line.


\section{X gate}\label{Sec:x}

\subsection{Narrowband}\label{Sec:x-nb}
Derivation of the narrowband (hence passband) CPs, contrary to the broadband ones, reveals that they must have asymmetric design of phases, which puts up a barrier for derivation of longer sequences due to heavy numerical calculations. As it was mentioned, narrowband pulses are prone to superfluous wiggles when derived by SU(2) method. Nevertheless, we derive composite X gates by this method. To accomplish that, we set two appropriate designs of CPs --- antisymmetric A$N$ and Wimperis-kind W$N$, both are the sequences of $\pi$ pulses. 

If we target pure $\pi$ composite rotations ($\phi = 0$), A$N$ has the following structure or design 
\be\label{X-AN}
\pi_{\phi_1} \pi_{\phi_2} \cdots \pi_{\phi_{n_s}} \pi_{\phi_{n_s + 1}} \pi_{-\phi_{n_s}} \cdots \pi_{-\phi_2} \pi_{-\phi_1},
\ee
and consists of the odd number of $\pi$ pulses, which besides the middle one, have phases with equal absolute value but with opposite signs when tracking from the left to right and from the right to left. Since we target X gate ($\phi = \pi/2$), $\pi/2$ is added to all phases with both minus and plus signs $\pm \phi_k \rightarrow \pm \phi_k + \pi/2$.

Again, for $\pi$ composite rotations ($\phi = 0$), W$N$ design looks more interpretable 
\be\label{X-WN}
\pi_{\phi_1} \pi_{\phi_2} \pi_{\phi_3} \cdots \pi_{\phi_{n_s + 1}} \pi_{\phi_{n_s + 1}} \cdots \pi_{\phi_3} \pi_{\phi_2},
\ee
and consists of the odd number of $\pi$ pulses, where, besides the first pulse, the second half of the structure is a mirror image of the first half, i.e., in the second half, phases are written in the opposite direction. Again, since we target X gate ($\phi = \pi/2$), $\pi/2$ is added to all phases $\phi_k \rightarrow \phi_k + \pi/2$. 

Interestingly, one gets wiggles in the case of A$N$ with 5, 9, 13 pulses, but the gap was filled with W$N$, which are useful for 5, 9, 13, ... . Lowest member of W$N$ is the well-known NB1 pulse of Wimperis (W$5$), hence the name of the design. A$N$ and W$N$ CPs for X gate derived by SU(2) approach are listed in Table~\ref{Table:NB-X-AN}. We choose FWHM as the performance measure of sensitivity (narrowness) of narrowband composite X gates.

\begin{figure}[t]
\bt{r}
\centerline{\includegraphics[width=0.8\columnwidth]{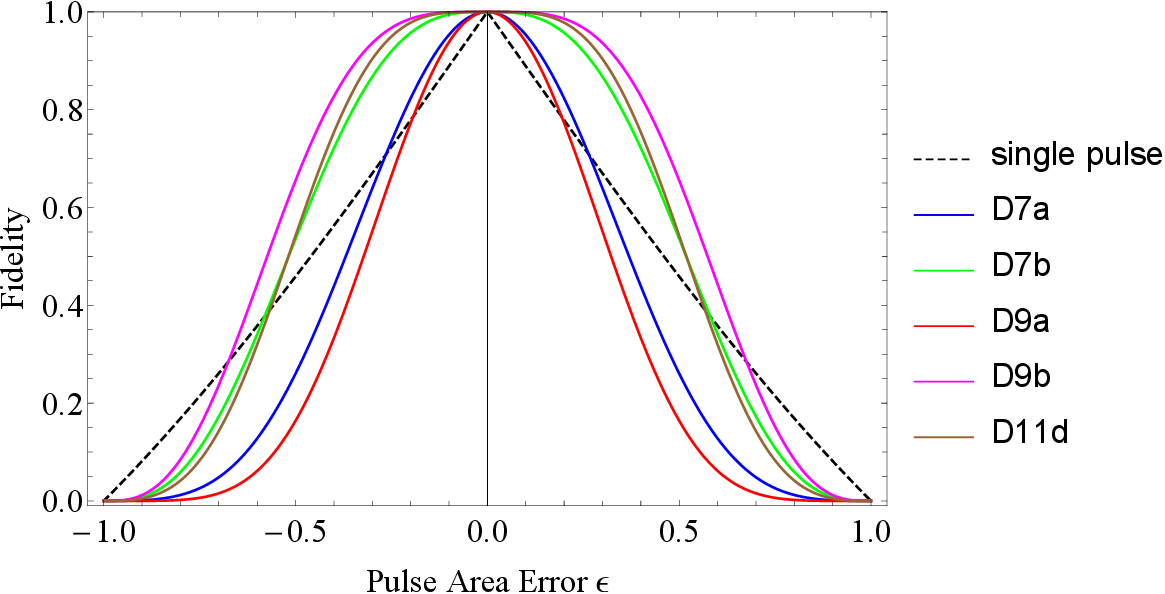}} \\ \\
\centerline{\includegraphics[width=0.8\columnwidth]{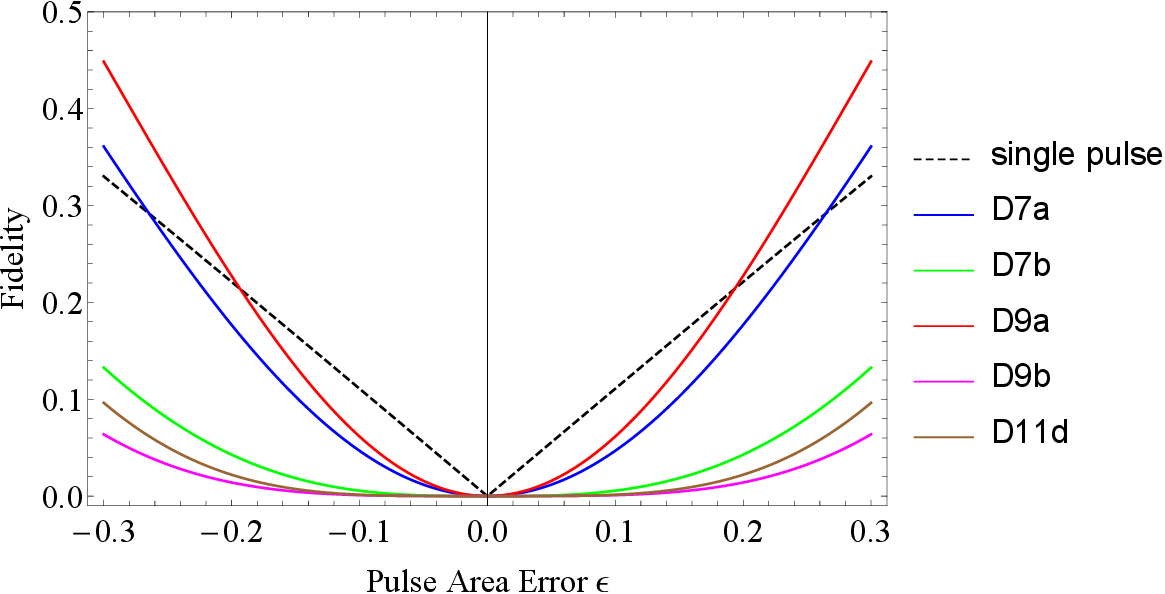}}
\et
\caption{
Frobenius distance fidelity (top) and infidelity (bottom) of composite passband X gates produced by D$N$ (diversis passuum) sequences from the Table~\ref{Table:PB-X-DN}.
}
\label{fig:PB-X-DN}
\end{figure}

Curiously, for X gate or $\pi$ rotation, the modified-SU(2) approach improves the results obtained by the SU(2) method. We derived up to 13 CPs by this method, called A$N$-m, which have the same antisymmetric design of A$N$. For example, A$3$ is derived by both methods. Table~\ref{Table:NB-X-ANm} shows that the A$N$-m CPs for X gate derived by the modified version of SU(2) outperform the same members $N$ obtained by conventional one. This shows the privilege of antisymmetric pulses over Wimperis-kind ones.

\subsection{Passband}\label{Sec:x-pb}
\subsubsection{Pari passu}
Difficulty of derivation of the passband rotation gates is manifested in the appearance of alternation in fidelity (deriving by SU(2)), not established by the derivation method, exhibiting their tenderness. This problem can be solved by using a regularization method, instead of the strict SU(2) method. Despite that, it was possible to derive SK1 and PB1 as the first and second order pari passu passband pulses, respectively (wiggles arise for longer sequences). In both methods, the design of pari passu passband pulses is the same
\be\label{X-P}
\pi_{\phi_1} (2\pi)_{\phi_2} (2\pi)_{\phi_3} \cdots (2\pi)_{\phi_{N}},
\ee
the sequence of nominal $2\pi$ pulses, preceded (or succeeded) by a
pulse of area $\pi$; and the number of pulses $N$ is odd. Sequences obtained by regularization method P$N$ for X gate are listed in Table~\ref{Table:PB-X-PN}. The first member P3 is SK1 ($n_p = 1$), and the second member P5 outperforms PB1 (both $n_p = 2$) by means of error sensitivity range and rectangularity. Increasing the number of pulses, the performance measures, namely, sensitivity, robustness and rectangularity, improve regularly. This is not the case for SU(2) method, when using ultrahigh-precision measures --- the UL-fidelity range of PB1 remains equal to the same of SK1.

Intersection of fidelities of P3 and P5 is at $\epsilon_{35} = 0.454371$ and $\mathcal{F}(\epsilon_{35}) = 0.461157$; accordingly for P5 and P7: $\epsilon_{57} = 0.471023$ and $\mathcal{F}(\epsilon_{57}) = 0.423852$; and for P7 and P9: $\epsilon_{79} = 0.478761$ and $\mathcal{F}(\epsilon_{79}) = 0.403605$.
It seems that intersection $\epsilon_{N-2,N}$ converges to $\epsilon_p = 0.5$ when $N \gg 1$, and one may get square fidelity for sufficiently large $N \geq N_p$. To obtain $N_p$ seems to be done by supercomputer (or maybe quantum computer), neural networks, or their combination, depending on the complexity of the optimization algorithm. 

In our opinion, the value $\epsilon_p = 0.5$ is suggested by the method we use, since the fidelity both at the bottom and at the top is optimized with equal force and with equal step (pari passu). 

\subsubsection{Diversis passuum}
Heterogeneous optimization of broadband and narrowband properties generates another type of passband pulses, called diversis passuum, which can be derived using SU(2) method, denoted as D$N$ 
\be\label{X-D}
\pi_{\phi_1} \pi_{\phi_2} \cdots \pi_{\phi_{N}},
\ee
which don't have a special design in general, although for the lowest members D$7a$ and D$7b$ phases have a simple structure (see Table~\ref{Table:PB-X-DN}).

\begin{figure}[t]
\bt{r}
\centerline{\includegraphics[width=0.8\columnwidth]{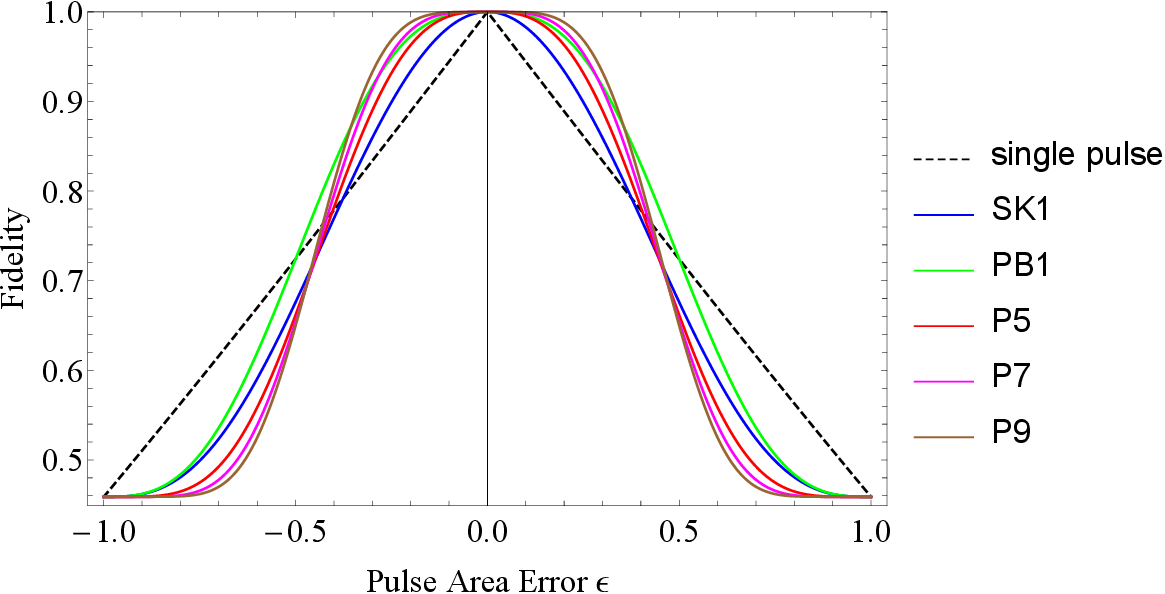}} \\ \\
\centerline{\includegraphics[width=0.8\columnwidth]{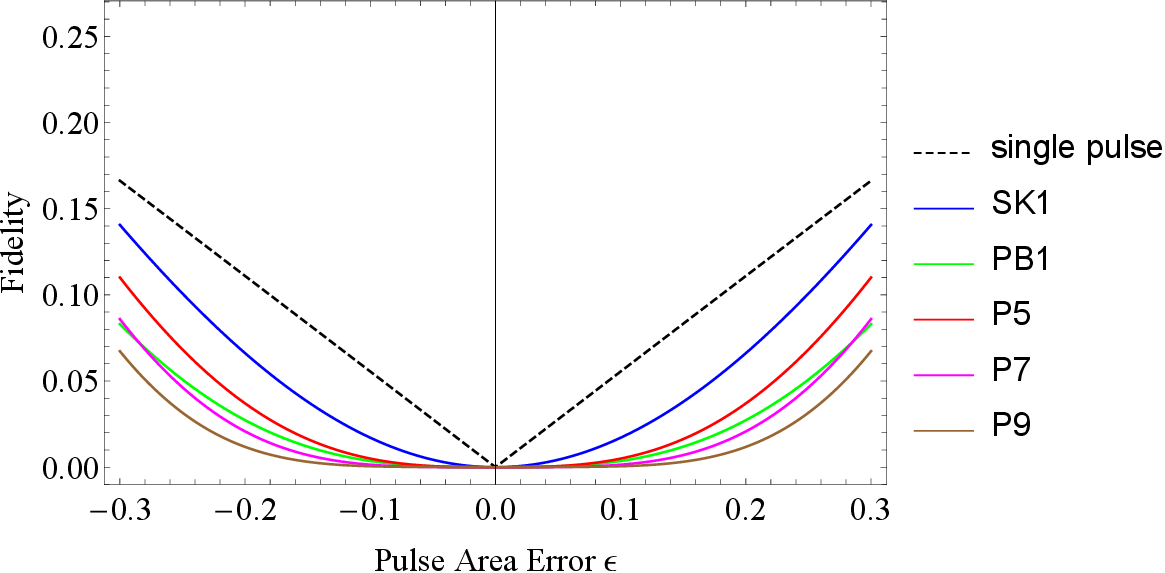}}
\et
\caption{
Frobenius distance fidelity (top) and infidelity (bottom) of composite passband Hadamard gates produced by P$N$ (pari passu) sequences from the Table~\ref{Table:PB-H-PN}.
}
\label{fig:PB-H-PN}
\end{figure}


\section{Hadamard gate}\label{Sec:h}
\subsection{Narrowband}\label{Sec:h-nb}
In the case of non-$\pi$ rotations, both major and minor element dependences $\mathcal{U}_{11}(\epsilon)$ and $\mathcal{U}_{21}(\epsilon)$ are asymmetric and can not be modified to be symmetric. Considering also the fact that the useful CPs are asymmetric for narrowband rotations, the hard numerical calculations are necessary for obtaining the results. 

For optimization of non-$\pi$ rotations, the SU(2) method is used. Four designs or structures can be used to derive narrowband Hadamard gate --- antisymmetric 1st type and 2nd type, Wimperis-kind and asymmetric, general structure of which liberally presented in Sec.~\ref{Sec:gen-nb}. Corresponding members of these four families are displayed in Table~\ref{Table:NB-H} and Fig.~\ref{fig:NB-H}.

Two facts must be acknowledged for rotations other than $\pi$: 

\begin{itemize}
\item The error sensitivity range of the resulting narrowband CPs depends on the design (the structure of the pulse areas) used rather than the order of optimization. For a particular structure, increasing the optimization order for longer pulses makes the pulses narrower, but the different structures differ in error sensitivity ranges for the same optimization order. The reason is that different structures already optimize the overall gate error at different levels. For example, the structure A$N$ without optimization already nullifies the odd derivatives $1, 3, 5, 7, \ldots$ of $\, \mathcal{U}_{11}(\epsilon)$ at $\epsilon = - 1$ and the even derivatives $2, 4, 6, 8, \ldots$ of $\, \mathcal{U}_{21}(\epsilon)$ at $\epsilon = - 1$. 
\item The case of X gate is special: the number of pulses of A$N$ is lower by two --- A5 of Hadamard gate (and non-$\pi$ general rotation gate) is equivalent to A3 of X gate, A7 to A5, and so forth. 
\end{itemize}

\begin{figure}[t]
\bt{r}
\centerline{\includegraphics[width=0.8\columnwidth]{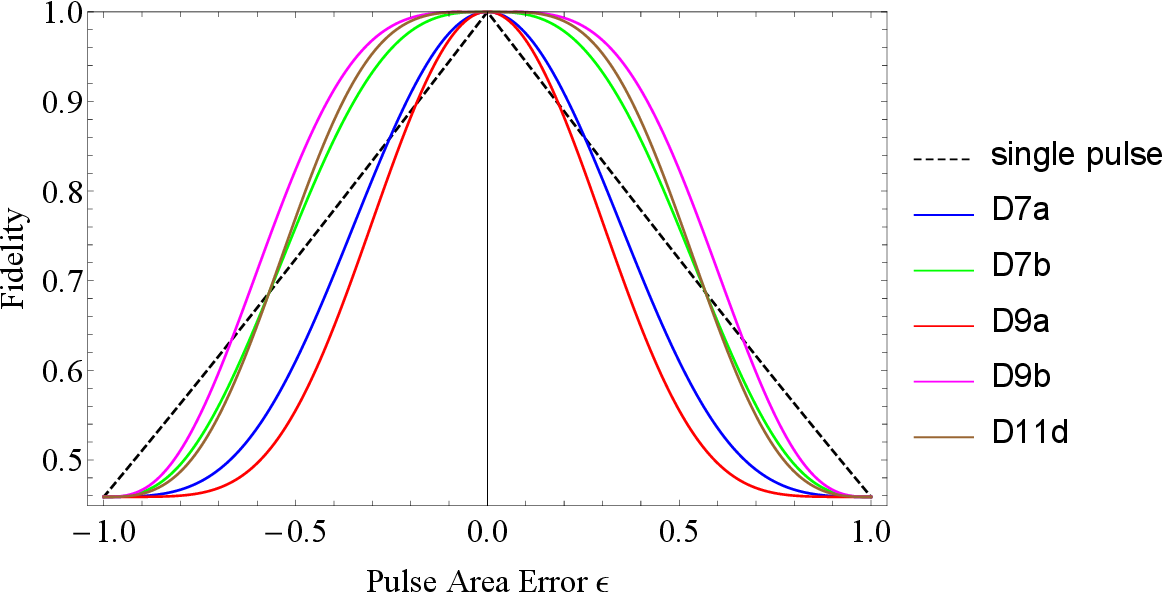}} \\ \\
\centerline{\includegraphics[width=0.8\columnwidth]{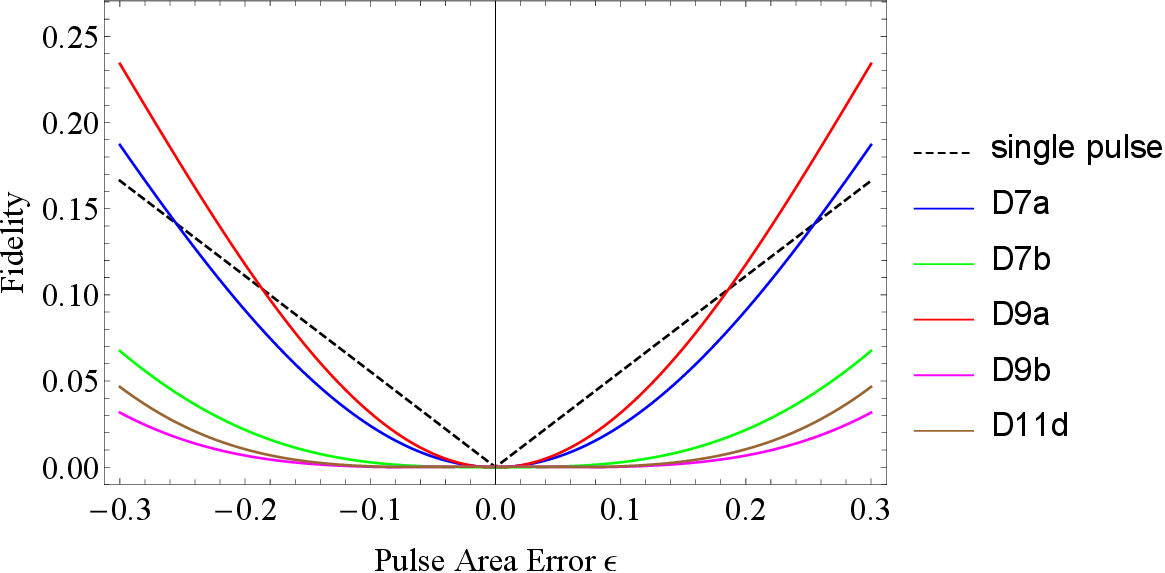}}
\et
\caption{
Frobenius distance fidelity (top) and infidelity (bottom) of composite passband Hadamard gates produced by D$N$ (diversis passuum) sequences from the Table~\ref{Table:PB-H-DN}.
}
\label{fig:PB-H-DN}
\end{figure}

\subsection{Passband}\label{Sec:h-pb}
Pari passu passband P$N$ CPs for Hadamard gate have the structure presented in Sec.~\ref{Sec:gen-pb} and are displayed in Table~\ref{Table:PB-H-PN}.

Intersection of fidelities of P3 and P5 is at $\epsilon_{35} = 0.44608$ and $\mathcal{F}(\epsilon_{35}) = 0.726128$; accordingly for P5 and P7: $\epsilon_{57} = 0.465489$ and $\mathcal{F}(\epsilon_{57}) = 0.702619$; and for P7 and P9: $\epsilon_{79} = 0.474985$ and $\mathcal{F}(\epsilon_{79}) = 0.689117$.
As for X gate, here also we see tendency $\epsilon_{N-2,N} \rightarrow \epsilon_{p}$ when $N \gg 1$.

Diversis passuum passband D$N$ CPs for Hadamard gate have the structure presented in Sec.~\ref{Sec:gen-pb} and are displayed in Table~\ref{Table:PB-H-DN}.

\section{General rotation gate}\label{Sec:gen}
General rotation gates, narrowband and passband, being non-$\pi$ rotations, can be obtained in the same fashion as Hadamard gate ($\theta = \pi/2$).

\subsection{Narrowband}\label{Sec:gen-nb}
Generalization of A$N$ CPs, i.e., an antisymmetric sequencese of 1st type, have the following structure, in general, presented by $\theta$ parameter
\be\label{G-AN}
\left({\frac{\pi - \theta}{2}}\right)_{\phi_0} \pi_{\phi_1} \cdots \pi_{\phi_{n_s}} \pi_{\phi_{n_s + 1}} \pi_{-\phi_{n_s}} \cdots \pi_{-\phi_1} \left({\frac{\pi - \theta}{2}}\right)_{-\phi_0}.
\ee
When targeting general rotation gates ($\phi = \pi/2$), as usual, this $\pi/2$ phase change must be done for all the components in the structure. For non-$\pi$ rotations, the number of pulses is $N = 2 n_s + 3$, where $n_s$ is the sensitivity order. In the case of $\pi$ rotations ($\theta = \pi$), we transition to the Eq.~\eqref{X-AN}, where one gets rid of the first and the last pulses (being zero rotations), hence $\phi_0$, and the number of constituent pulses becomes $N = 2 n_s + 1$. General formula for the number of pulses and total operation time can be presented using a step function $\sigma$,
\bse\label{T-G-AN}
\begin{align}
& N(\theta) = 2 n_s + 1 + 2 \sigma(\theta), \\
& \mathcal{A}_{tot}(\theta) = N(\theta) \pi - 2 \theta, \\
& \sigma(\theta) = 
    \begin{cases}
        1 & \text{if } \theta \in (0, \pi), \\
        0 & \text{if } \theta = \pi.
    \end{cases}
\end{align}
\ese

In the case of the 1st type, we pre-set the structure \eqref{G-AN} and optimization is done afterwards. As was mentioned, the performance of the resulting narrowband CPs depends on the used design rather than the order of optimization. Already, choosing the design, one may get many derivatives equal to zero beforehand. This is a case for the 1st type. Although of some lose on speed of operation, A$N$ is much robust and has systematic design compared to the rest, providing systematic pattern in the performance measures. For the same $n_s$ it gives the best performance. For $\pi$ rotations A$N$-m are the best ones having the same A$N$ design. 

Alternatively, one may use 2nd type of antisymmetric design AT$N$
\be\label{G-ATN}
\alpha_{\phi_1} \pi_{\phi_2} \cdots \pi_{\phi_{n_s+1}} \pi_{-\phi_{n_s+1}} \cdots \pi_{-\phi_2} \alpha_{-\phi_1},
\ee
where to the overall phase structure the $\pi/2$ phase is added, to obtain a general rotation gate. When $\theta = \pi$, $\alpha$ is equal to $\pi/2$ and one gets the alternative to \eqref{X-AN} structure which have the same performance (in $\mathcal{A}_{tot}$ and sensitivity). Results of 2nd type with the number $N+1$ derived by SU(2) method will not differ in either speed or sensitivity from the 1st type with the number $N$ in Table~\ref{Table:NB-X-AN}. So, the separation of the antisymmetric sequences to two types becomes important in the case of non-$\pi$ rotation gates.

The two asymmetric sequences are also useful for general rotation gates --- Wimperis-kind and ``just'' asymmetric. Wimperis-kind W$N$ design, written for $\phi = 0$, is
\be\label{G-WN}
\theta_{\phi_1} \pi_{\phi_2} \cdots \pi_{\phi_{2 n_s + 1}},
\ee
phases of which have simpler structure in the case of $N = 5, 9, 13, \ldots$:
\be\label{G-WN-a}
\theta_{\phi_1} \pi_{\phi_2} \cdots \pi_{\phi_{n_s + 1}} \pi_{\phi_{n_s + 1}} \pi_{\phi_{n_s}} \cdots \pi_{\phi_{2}}.
\ee
Unfortunately, W$N$ didn't have 1st order member. The lowest member is the 2nd order pulse known as NB1 with the simplified structure (again for $\phi = 0$)
\be\label{G-NB1}
\theta_{0} \pi_{\chi} \pi_{- \chi} \pi_{- \chi} \pi_{- \chi},
\ee
where $\chi = \arccos \left( - \frac{\theta}{4\pi} \right)$. To obtain rotation gates from these sequences, one must add $\pi/2$ to the phase structures in~\eqref{G-WN},~\eqref{G-WN-a} and~\eqref{G-NB1}. For X gate $\chi \approx 0.580431\pi$ and for Hadamard $\chi \approx 0.539893\pi$.

Since the sequence of $\pi$ pulses in the CP carry the optimization process and seeds a stable design, the most fictitious asymmetric pulse may have the following design
\be\label{G-ASN}
\alpha_{\phi_1} \pi_{\phi_2} \pi_{\phi_3} \cdots \pi_{\phi_{2 n_s-1}} \pi_{\phi_{2 n_s}} \beta_{\phi_{2 n_s +1}},
\ee
denoted as AS$N$. Sometimes, it is possible to find the best trade-off in speed and accuracy by these sequences. Good example is AS$9$ in Table~\ref{Table:NB-H}.

Both Wimperis-kind and asymmetric designs converge to the sequence of $\pi$ pulses in the case of $\pi$ rotations (X gate), and, in that case, the most neat structure is antisymmetric, i.e. A$N$.

\subsection{Passband}\label{Sec:gen-pb}
Pari passu passband rotation gates P$N$ are subjected to the following design ($\phi = \pi/2$ must be added to all the phases)
\be\label{G-PN}
\theta_{\phi_1} (2\pi)_{\phi_2} (2\pi)_{\phi_3} \cdots (2\pi)_{\phi_{2n_p +1}},
\ee
which can be considered as the generalization of the SK1: 
\be\label{G-SK1}
\theta_{0} (2\pi)_{\chi} (2\pi)_{-\chi},
\ee
where $\chi = \arccos \left( - \frac{\theta}{4\pi} \right)$, and the PB1:
\be\label{G-PB1}
\theta_{0} (2\pi)_{\chi} (2\pi)_{-\chi} (2\pi)_{-\chi} (2\pi)_{\chi},
\ee
where $\chi = \arccos \left( - \frac{\theta}{8\pi} \right)$. 

Both SK1 and PB1 can be derived using SU(2) method, but, unfortunately, this optimization is strict to use for longer sequences. Longer ones can be obtained using another propagator-optimization method, where the fidelity is optimized instead of matrix elements. It also ensures constant (at the center) and sensitive (on the edges) rotations due to regularizer used, hence, we call it a regularization method. Both methods can be used to derive any $\theta$ rotation gate.  

Diversis passuum passband rotation gates D$N$ have the design similar to~\eqref{G-WN} 
\be\label{G-DN}
\theta_{\phi_1} \pi_{\phi_2} \cdots \pi_{\phi_{2 (n_r + n_s) + 1}},
\ee
but here the number of pulses is equal to $N = 2 (n_r + n_s) + 1$. It was possible to find the simpler structure of phases for the lowest members: more sensitive D$7a$ with $(n_s, n_r) = (2, 1)$ and more robust D$7b$ with $(n_s, n_r) = (1, 2)$ 
\be\label{G-D7}
\theta_{\phi_1} \pi_{\phi_2} \pi_{\phi_3} \pi_{\phi_4} \pi_{-\phi_3} \pi_{-\phi_4} \pi_{\phi_2 - \pi}.
\ee

Although we use only SU(2) method to derive D$N$-s, the regularization method can be applied as an alternative.

\section{Conclusions\label{Sec:concl}}

We presented CPs which produce narrowband and passband rotational single-qubit gates, namely --- X, Hadamard and general rotation gates. Narrowband and passband CPs tend to alternate (wiggle) in fidelity, which is an unintended result of derivation method due to the nature of the problem. Furthermore, having the same order of optimization, these CPs differ in performance depending on the structure of the pulse area (design) and the method of derivation used.

Three types of optimization methods are used --- SU(2), modified-SU(2), and regularization. Narrowband X gates, derived by the modified-SU(2) approach, are superior to the corresponding gates, obtained using the SU(2) approach by the means of sensitivity, for the same sensitivity order. For example, the antisymmetric A5-m pulse outperforms the well-known NB1 --- FWHM of A5-m is about 42.8\%, which is narrower than FWHM of NB1 49.4\% of the whole error bandwidth. Since we have not found an alternative to the SU(2) method for narrowband Hadamard or general rotation gates, we apply this old method to two antisymmetric and two asymmetric pulse designs.

We propose the two types of passband CPs --- \emph{pari passu} P$N$, with passband order, and \emph{diversis passuum} D$N$, with different sensitivity and robustness orders. P$N$ sequences are derived by the regularization method, and show systematic improvement in all the performance characteristics --- sensitivity, robustness and rectangularity. D$N$ sequences are derived by the SU(2) method, although regularization method can also be used.

The results in this chapter can be useful in applications such as spatial localization in \emph{in vivo} NMR spectroscopy, selective and local spatial addressing of trapped ions or atoms in optical lattices by tightly focused laser beams in quantum sensing, narrowband polarization filters and passband polarization retarders in polarization optics.




\acknowledgments
H.L.G. acknowledges support from the EU Horizon-2020 ITN project LIMQUET \textit{Light-Matter Interfaces for Quantum Enhanced Technologies} (Contract No. 765075) and from the Higher Education and Science Committee of Armenia in the framework of the research project 21AG-1C038 on \textit{Methods of Information Theory in Statistical Physics and Data Science}.


\appendix

\section{The form of the composite rotation gate and it's parameters}

Considering a pulse area error $\epsilon$, the errant (actual) propagator can be represented in the following way 
\be
\boldsymbol{\mathcal{U}}(\epsilon) = \left[ \begin{array}{cc} e^{i \gamma (\epsilon)} \cos(\A (\epsilon) /2) & 
 - i e^{i\phi(\epsilon)} \sin(\A (\epsilon) /2) \\ - i e^{- i\phi(\epsilon)} \sin(\A (\epsilon) /2) & e^{- i \gamma (\epsilon)} \cos(\A (\epsilon) /2) \end{array} \right], \nonumber 
\ee
where now an overall geometrical phase $\gamma(\epsilon)$ and an overall gate phase [which is stable/robust in the case of constant rotations (the class of broadband CPs), and sensitive (and robust) in the case of our narrowband (passband) CPs] $\phi(\epsilon)$ are errant. These and an overall pulse area $\A (\epsilon)$ error dependences are nonlinear, in general, for composite rotations. Our target are the quantum $Y$ rotation gates $[\A_{(\epsilon = 0)} = \theta$, $\phi_{(\epsilon = 0)} = \pi/2$ and $\gamma_{(\epsilon = 0)} = 0]$. Then, one can easily find that, nullifying the derivatives (Taylor coefficients) of major and minor diagonal elements, the derivatives of $\phi(\epsilon)$ and $\A (\epsilon)$ will be also nullified. Hence, one can have sensitivity or/and robustness of these parameters.  

\section{Tables of the composite phases and the precision measures of the composite pulses}

Here we present the complete sets of phases of the composite pulse sequences.

\begin{table*}[htp]
\centering
\begin{tabular}{ccclc}
\hline
\hline
\\
Name & Pulses & $O(\epsilon^{n_s})$ & Phases (in units $\pi$) & FWHM (error sensitivity range) \\
\\
\hline
\\
     &        &                 & Antisymmetric sequences             & \\
     &        &                 & $\{\phi_1, \phi_2, \ldots, \phi_{n_s}, \phi_{n_s+1}, - \phi_{n_s}, \ldots, - \phi_2, - \phi_1\}+1/2$ & \\        
single & 1 & $O(\epsilon^0)$ & $0$ & $ [0.53989\pi, 1.46011\pi] $ \\
A3 & 3 & $O(\epsilon)$ & $\frac13, 1$ & $ [0.726\pi, 1.274\pi] $ \\
A7 & 7 & $O(\epsilon^3)$ & $0.244, 1.6719, 0.7626, 1$ & $ [0.802\pi, 1.198\pi] $ \\
A11 & 11 & $O(\epsilon^5)$ & $0.3468, 1.0836, 0.6708, 0.8186, 0.2655, 1$ & $ [0.823\pi, 1.177\pi] $ \\
\\
\hline
\\
     &        &                 & Wimperis-kind sequences             & \\
     &        &                 & $\{\phi_1, \phi_2, \ldots, \phi_{n_s-1}, \phi_{n_s+1}, \phi_{n_s+1}, \phi_{n_s-1}, \ldots, \phi_2\}+1/2$ & \\  
W5 $\overset{\Delta}{=}$ NB1 & 5 & $O(\epsilon^2)$ & $0, 0.5804, 1.4196$ & $ [0.753\pi, 1.247\pi] $ \\    
W9 & 9 & $O(\epsilon^4)$ & $0, 0.4417, 1.8105, 0.8737, 1.3$ & $ [0.813\pi, 0.187\pi] $ \\  
W13 & 13 & $O(\epsilon^6)$ & $0, 1.2863, 0.7778, 1.4773, 0.3446, 0.5632, 1.7414$ & $ [0.861\pi, 1.139\pi] $ \\
\\
\hline
\hline
\end{tabular}
\caption{
Phases of antisymmetric composite sequences of $N=2n_s+1$ nominal $\pi$ pulses, which produce the $\pi$ rotation with a pulse area error sensitivity up to order $O(\epsilon^{n_s})$.
The last column gives the half-fidelity range or, so called, FWHM (full width at half maximum) $[\pi (1-\epsilon_0), \pi (1+\epsilon_0)]$ of pulse area error sensitivity wherein the Frobenius distance fidelity is above the value $0.5$, i.e. the infidelity is below HM$=\frac{MAX + MIN}{2} = \frac{1 + 0}{2} = 0.5$. \\
}
\label{Table:NB-X-AN}
\end{table*}

\begin{table*}[htp]
\centering
\begin{tabular}{ccclc}
\hline
\hline
\\
Name & Pulses & $O(\epsilon^{n_s})$ & Phases (in units $\pi$) & FWHM \\
& & & & (error sensitivity range) \\
\\
\hline
\\
     &        &                 & Antisymmetric sequences             & \\
     &        &                 & $\{\phi_1, \phi_2, \ldots, \phi_{n_s}, \phi_{n_s+1}, -\phi_{n_s}, \ldots, -\phi_2, -\phi_1\}+1/2$ & \\        
single & 1 & $O(\epsilon^0)$ & $0$ & $ [0.53989\pi, 1.46011\pi] $ \\
A3 & 3 & $O(\epsilon)$ & $\frac13, 1$ & $ [0.726\pi, 1.274\pi] $ \\
A5-m & 5 & $O(\epsilon^2)$ & $\frac{4}{5}, \frac{8}{5}, 0$ & $ [0.786\pi, 1.214\pi] $ \\
 &  &  & $\frac{6}{5}, \frac{2}{5}, 0$ &  \\
A7-m & 7 & $O(\epsilon^3)$ & $\frac{5}{7}, \frac{13}{7}, \frac{11}{7}, 1$ & $ [0.819\pi, 1.181\pi] $ \\
 &  &  & $\frac{9}{7}, \frac{1}{7}, \frac{3}{7}, 1$ &  \\
A9-m & 9 & $O(\epsilon^4)$ & $\frac{2}{9}, \frac{12}{9}, \frac{10}{9}, \frac{4}{9}, 0$ & $ [0.840\pi, 1.160\pi] $ \\
 & & & $\frac{14}{9}, \frac{12}{9}, \frac{16}{9}, \frac{10}{9}, 0$ & \\
A11-m & 11 & $O(\epsilon^5)$ & $\frac{7}{11}, \frac{1}{11}, \frac{13}{11}, \frac{17}{11}, \frac{19}{11}, 1$ & $ [0.855\pi, 1.145\pi] $ \\
 & & & $\frac{9}{11}, \frac{17}{11}, \frac{1}{11}, \frac{3}{11}, \frac{15}{11}, 1$ & \\
A13-m & 13 & $O(\epsilon^6)$ & $\frac{20}{13}, \frac{18}{13}, \frac{22}{13}, \frac{16}{13}, \frac{24}{13}, \frac{14}{13}, 0$ & $ [0.867\pi, 1.133\pi] $ \\
\\
\hline
\hline
\end{tabular}
\caption{
Phases of antisymmetric composite sequences of $N=2n_s+1$ nominal $\pi$ pulses, which produce the $\pi$ rotation with a pulse area error sensitivity up to order $O(\epsilon^{n_s})$.
The last column gives the half-fidelity range or, so called, FWHM (full width at half maximum) $[\pi (1-\epsilon_0), \pi (1+\epsilon_0)]$ of pulse area error sensitivity wherein the Frobenius distance fidelity is above the value $0.5$, i.e. the infidelity is below HM$=\frac{MAX + MIN}{2} = \frac{1 + 0}{2} = 0.5$. \\
}
\label{Table:NB-X-ANm}
\end{table*}

\begin{table*}[htp]
\centering
\begin{tabular}{ccclcc}
\hline
\hline
\\
Name & Pulses & $O(\epsilon^{n_s})$ & Phases (in units $\pi$) & $\mathcal{A}_{\text{tot}}$ & FWHM \\
 & & & & & (error sensitivity range) \\
\\
\hline
\\
     &        &                 & Antisymmetric sequences (1st type)           & & \\
     &        &                 & $\{\phi_0, \phi_1, \phi_2, \ldots, \phi_{n_s}, \phi_{n_s+1}, - \phi_{n_s}, \ldots, - \phi_2, - \phi_1, -\phi_0 \}+1/2$  & & \\
single & 1 & $O(\epsilon^0)$ & $0$ & $\pi/2$ & $ [0.50973\pi, 1.49027\pi] $ \\
A5 & 5 & $O(\epsilon)$ & $1, 0.2301, 1$ & $3.5\pi$ & $ [0.725\pi, 1.275\pi] $ \\
A7 & 7 & $O(\epsilon^2)$ & $1, 0.2954, 0.8230, 0$ & $5.5\pi$ & $ [0.7987\pi, 1.2013\pi] $ \\
A9 & 9 & $O(\epsilon^3)$ & $1, 0.3082, 0.7709, 0.1152, 1$ & $7.5\pi$ & $ [0.8123\pi, 1.1877\pi] $ \\
A11 & 11 & $O(\epsilon^4)$ & $1, 1.9962, 0.8077, 1.4886, 0.6279, 0$ & $9.5\pi$ & $ [0.8006\pi, 1.1994\pi] $ \\
A13  & 13 & $O(\epsilon^5)$ & $1, 0.2291, 1.9036, 1.5659, 0.5577, 0.9389, 1$ & $11.5\pi$ & $ [0.8306\pi, 1.1694\pi] $ \\
\hline
\\
     &        &                 & Asymmetric sequences (Wimperis-kind)           & & \\
     &        &                 & $\{\phi_1, \phi_2, \ldots, \phi_{2 n_s + 1} \}+1/2$ & & \\           
W5 $\overset{\Delta}{=}$ NB1 & 5 & $O(\epsilon^2)$ & $0, 0.539893, -0.539893, -0.539893, 0.539893$ & $4.5\pi$ & $ [0.7396\pi, 0.2604\pi] $ \\
W7 & 7 & $O(\epsilon^3)$ & $0, 0.958038, 1.70099, 1.13518, 1.91065, 0.793721, 0.27538$ & $6.5\pi$ & $ [0.7897\pi, 1.2103\pi] $ \\
W9 & 9 & $O(\epsilon^4)$ & $0, 1.26578, 0.415523, 0.152784, 1.2404, 1.2404, 0.152784,$ & $8.5\pi$ & $ [0.8105\pi, 1.1895\pi] $ \\
 & & & $0.415523, 1.26578$ & & \\
W13  & 13 & $O(\epsilon^6)$ & $0, 1.88926, 0.801439, 1.32824, 0.606488, 1.39505, 0.211373, 0.211373$ & $12.5\pi$ & $ [0.8332\pi, 1.1668\pi] $ \\
 &  &  & $1.39505, 0.606488, 1.32824, 0.801439, 1.88926$ & &  \\
\hline
\hline
\\
     &        &                 & Asymmetric sequences            & & \\
     &        &                 & ${\alpha}_{\phi_1} \pi_{\phi_2} \cdots \pi_{\phi_{N-1}} \beta_{\phi_{N}}$ & & \\   
     &        &                 & $\alpha, \beta; \phi_1, \phi_2, \ldots, \phi_N $ & & \\ 
AS3 & 3 & $O(\epsilon)$ & $0.486, 1.2463; 0.2540, 1.8598, 0.9799$ & $2.7323\pi$ & $ [0.7204\pi, 1.2796\pi] $ \\ 
AS5 & 5 & $O(\epsilon^2)$ & $0.6298, 0.8426; 0.4083, 1.6092, 0.7888, 1.4389, 0.3147$ & $4.4723\pi$ & $ [0.7407\pi, 1.2593\pi] $ \\
AS5a & 5 & $O(\epsilon^2)$ & $0.778276, 0.69444; 1.8521, 0.997442, 0.280465, 1.05988, 1.88585$ & $4.4727\pi$ & $ [0.741\pi, 0.259\pi] $ \\
AS7 & 7 & $O(\epsilon^3)$ & $0.38093, 1.2554; 1.7684, 1.9270, 0.8589, 1.3447, 0.6742, 1.4987, 0.3302$ & $6.6363\pi$ & $ [0.797\pi, 1.203\pi] $ \\
AS7a & 7 & $O(\epsilon^3)$ & $0.320885, 1.31241; 0.27189, 0.17642, 1.26598, 0.642382,$ & $6.6333\pi$ & $ [0.797\pi, 1.203\pi] $ \\
 &  &  & $1.33723, 0.600511, 1.72175$ &  &  \\
AS9 & 9 & $O(\epsilon^4)$ & $0.33322, 0.21161; 1.13696, 1.70097, 1.57736, 0.33491, 1.46716,$ & $7.5448\pi$ & $ [0.8314\pi, 1.1686\pi] $ \\
 &  &  & $0.61094, 0.87834, 0.32587, 1.19906$ &  & \\
\hline
\\
     &        &                 & Antisymmetric sequences (2nd type)          & & \\
     &        &                 & $\alpha; \{\phi_1, \phi_2, \ldots, \phi_{n_s+1}, - \phi_{n_s+1}, \ldots, - \phi_2, - \phi_1\}+1/2$ & & \\     
AT4 & 4 & $O(\epsilon)$ & $0.344509; 0.32165, 0.55861$ & $2.689\pi$ & $ [0.7202\pi, 1.2798\pi] $ \\
AT6\footnote{AT6 is not flat bottom!} & 6 & $O(\epsilon^2)$ & $0.415106; 0.411914, 0.561681, 0.474778$ & $4.83\pi$ & $ [0.8243\pi, 1.1757\pi] $ \\
 & & & & & \\
AT6r\footnote{Regularization method is used for derivation. Although the compensation order (of the major and minor diagonal gate elements) is $O(\epsilon)$, the compensation of trace fidelity is actually of the order trace$O(\epsilon^3)$} & 6 & $O(\epsilon)$ & $0.382938; 0.374102, 0.560617, 0.486709$ & $4.77\pi$ & $ [0.7853\pi, 1.2147\pi] $ \\ 
AT8 & 8 & $O(\epsilon^3)$ & $0.317051; 0.275734, 0.230589, 1.334398, 0.651125$ & $6.634\pi$ & $ [0.7971\pi, 1.2029\pi] $ \\
AT10 & 10 & $O(\epsilon^4)$ & $0.255796; 0.085393, 1.667345, 0.759388, 1.0321, 1.927368$ & $8.512\pi$ & $ [0.8004\pi, 1.1996\pi] $ \\
AT12 & 12 & $O(\epsilon^5)$ & $0.307743; 0.2576, 0.157112, 1.53557, 0.46763, 1.25254, 0.70666$ & $10.615\pi$ & $ [0.8335\pi, 1.1665\pi] $ \\
AT14\footnote{AT14 is not perfectly flat bottom!} & 14 & $O(\epsilon^6)$ & $0.370181; 0.357786, 0.457094, 0.992936, 1.887852,$ & $12.74\pi$ & $ [0.8668\pi, 1.1332\pi] $ \\
 & & & $1.39542, 0.54164, 0.4294$ & & \\
 & & & & & \\
AT16 & 16 & $O(\epsilon^7)$ & $0.320743; 1.7175, 0.17708, 1.09267, 1.11819,$ & $14.641\pi$ & $ [0.8622\pi, 1.1378\pi] $ \\
 & & & $0.345712, 1.70626, 0.28594, 1.17393$ & & \\
\\
\hline
\hline
\end{tabular}
\caption{
Phases of A$N$: antisymmetric composite sequences of $N-2$ nominal $\pi$ pulses, sandwiched by two pulses of areas $\pi/4$, and W$N$: asymmetric Wimperis-kind composite sequences of $N-1$ nominal $\pi$ pulses, preceded (or succeeded) by a pulse of area $\theta = \pi/2$; which produce the $\theta = \pi/2$ rotation with a pulse area error sensitivity up to order $O(\epsilon^{n_s})$.
The last column gives FWHM (full width at half maximum) $[\pi (1-\epsilon_0), \pi (1+\epsilon_0)]$ of pulse area error sensitivity wherein the Frobenius distance fidelity is above the value $\frac{1+(1-\sqrt{1-1/\sqrt{2}})}{2}$, i.e. the infidelity is below HM$= \frac{MAX + MIN}{2} = \frac{1+(1-\sqrt{1-1/\sqrt{2}})}{2} \approx 0.7294$, where the minimum (at $\epsilon = \pm 1$) Frobenius distance fidelity for $\pi/2$ rotations is $MIN=1-\sqrt{1-1/\sqrt{2}} \approx 0.4588$. \\
}
\label{Table:NB-H}
\end{table*}

\begin{table*}[htp]
\centering
\begin{tabular}{ccclccc}
\hline
\hline
\\
Name & Pulses & $n_p$ & Phases (in units $\pi$) & UL-fidelity & UH-fidelity & $\Delta$\\
     &  &  & $\{\phi_1, \phi_2, \ldots, \phi_{N}\}+1/2$ & (error sensitivity range) & (error robustness range) & (rectangularity) \\
\\
\hline
\\  
single & 1 & $0$ & $0$ & $ [0.00013\pi, 1.99987\pi] $ & $ [0.99991\pi, 1.00009\pi] $ & $0.99978\pi$\\
\\  
P3 $\overset{\Delta}{=}$ SK1 & 3 & $1$ & $0, 0.58043, 1.41957$ & $ [0.027\pi, 1.973\pi] $ & $ [0.995\pi, 1.005\pi] $ & $0.967\pi$\\
P5 & 5 & $2$ & $0, 1.78506, 0.48331, 1.34363, 0.83030$ & $ [0.077\pi, 1.923\pi] $ & $ [0.978\pi, 1.022\pi] $ & $0.901\pi$\\
P7 & 7 & $3$ & $0, 1.07359, 0.772099, 1.32971, 0.436728,$ & $ [0.120\pi, 1.880\pi] $  & $ [0.957\pi, 1.043\pi] $ & $0.837\pi$\\
   & & & $1.69584, 0.0868198$ & & & \\
P9 & 9 & $4$ & $0, 1.97287, 0.139635, 1.67993, 0.430437,$ & $ [0.154\pi, 1.846\pi] $ & $ [0.936\pi, 1.064\pi] $ & $0.782\pi$\\
   & & & $1.33878, 0.762866, 1.11383, 0.97606$ & & & \\ 
\hline
\\
PB1 & 5 & $2$ & $0, 0.539893, 1.46011, 1.46011, 0.539893$ & $ [0.027\pi, 1.973\pi] $ & $ [0.976\pi, 1.024\pi] $ & $0.949\pi$\\
\\
\hline
\hline
\end{tabular}
\caption{
Phases of pari passu passband pulses for X gate. PB1-like pulse structure is used ($2\pi$ pulses with single $\pi$). \\
}
\label{Table:PB-X-PN}
\end{table*}

\begin{table*}[htp]
\centering
\begin{tabular}{ccclccc}
\hline
\hline
\\
Name & Pulses & $n_p$ & Phases (in units $\pi$) & UL-fidelity & UH-fidelity & $\Delta$\\
     &  &  & $\{\phi_1, \phi_2, \ldots, \phi_{N}\}+1/2$ & (error sensitivity range) & (error robustness range) & (rectangularity)\\
\\
\hline
\\  
single & 1 & $0$ & $0$ & $ [0.00019\pi, 1.99981\pi] $ & $ [0.99982\pi, 1.00018\pi] $ & $0.99963\pi$\\
\\     
P3 $\overset{\Delta}{=}$ SK1 & 3 & $1$ & $0, 0.539893, 1.46011$ & $ [0.031\pi, 1.969\pi] $ & $ [0.992\pi, 1.008\pi] $ & $0.961\pi$\\
P5 & 5 & $2$ & $0, 0.816729, 1.37808, 0.447255, 1.79388$ & $ [0.084\pi, 1.916\pi] $ & $ [0.973\pi, 1.027\pi] $ & $0.889\pi$\\
P7 & 7 & $3$ & $0, 1.07705, 0.752636, 1.35704, 0.410338,$ & $ [0.128\pi, 1.872\pi] $ & $ [0.949\pi, 1.051\pi] $ & $0.821\pi$\\
   & & & $1.71464, 0.0836752$ & & & \\
P9 & 9 & $4$ & $0, 0.85519, 0.665578, 1.75921, 1.44318,$ & $ [0.145\pi, 1.855\pi] $ & $ [0.947\pi, 1.053\pi] $ & $0.803\pi$\\
   & & & $0.526885, 0.761473, 1.67661, 1.84619$ & & & \\
\hline
\\
PB1 & 5 & $2$ & $0, 0.519907, 1.48009, 1.48009, 0.519907$ & $ [0.031\pi, 1.969\pi] $ & $ [0.970\pi, 1.030\pi] $ & $0.939\pi$\\
\\
\hline
\hline
\end{tabular}
\caption{
Phases of pari passu passband pulses for Hadamard gate. PB1-like pulse structure is used ($2\pi$ pulses with single $\pi/2$). \\
}
\label{Table:PB-H-PN}
\end{table*}

\begin{table*}[htp]
\centering
\begin{tabular}{ccclccc}
\hline
\hline
\\
Name & Pulses & $n_s, n_r$ & Phases (in units $\pi$) & UL-fidelity & UH-fidelity & $\Delta$\\
     &  &  & $\{\phi_1, \phi_2, \ldots, \phi_{N}\}+1/2$ & (error sensitivity range) & (error robustness range) & (rectangularity)\\
\\
\hline
\\  
     &        &                 & $\phi_1, \phi_2, \phi_3, \phi_4, -\phi_3, -\phi_4, \phi_2 - 1$ & & & \\      
D7a & 7 & $2, 1$ & $0, 1.24005, 0.851488, 0.371396$ & $ [0.043\pi, 1.957\pi] $ & $ [0.995\pi, 1.005\pi] $ & $0.952\pi$ \\
    & 7 & $2, 1$ & $0, 1.75995, 0.371396, 0.851488$ & & & \\
D7b & 7 & $1, 2$ & $0, 1.24005, 1.6286, 1.14851$ & $ [0.023\pi, 1.977\pi] $ & $ [0.974\pi, 1.026\pi] $ & $0.951\pi$ \\
    & 7 & $1, 2$ & $0, 0.759954, 0.371396, 0.851488$ & & & \\
\\
D9a & 9 & $3, 1$ & $0, 1.1196, 0.735454, 0.733506, 1.38161,$ & $ [0.109\pi, 1.891\pi] $ & $ [0.998\pi, 1.002\pi] $ & $0.889\pi$ \\
    & & & $0.680734, 1.93256, 1.56754, 0.0517938$ & & & \\
D9b & 9 & $1, 3$ & $0, 1.1196, 1.50375, 1.50181, 0.853711,$ & $ [0.020\pi, 1.980\pi] $ & $ [0.967\pi, 1.033\pi] $ & $0.947\pi$\\
    & & & $0.152812, 0.900982, 0.535975, 0.0517144$ & & & \\
 
D11d\footnote{Possible D11a, D11b and D11c, respectively, for $(n_s, n_r)=(4,1)$, $(1,4)$ and $(3,2)$, derived by our method, are alternating pulses with wiggles, not relevant to rotation gates.} & 11 & $2, 3$ & $0, 0.6661, 1.1385, 1.0056, 1.8812, 1.5879,$ & $ [0.028\pi, 1.972\pi] $ & $ [0.936\pi, 1.064\pi] $ & $0.908\pi$\\
    & & & $0.9755, 0.6004, 0.2092, 1.4173, 0.0729$ & & & \\
\\
\hline
\hline
\end{tabular}
\caption{
Phases of diversis passuum passband pulses for X gate. \\
}
\label{Table:PB-X-DN}
\end{table*}

\begin{table*}[htp]
\centering
\begin{tabular}{ccclccc}
\hline
\hline
\\
Name & Pulses & $n_s, n_r$ & Phases (in units $\pi$) & UL-fidelity & UH-fidelity & $\Delta$\\
     &  &  & $\{\phi_1, \phi_2, \ldots, \phi_{N}\}+1/2$ & (error sensitivity range) & (error robustness range) & (rectangularity)\\
\\
\hline
\\  
     &        &                 & $\phi_1, \phi_2, \phi_3, \phi_4, -\phi_3, -\phi_4, \phi_2 - 1$ & & & \\      
D7a & 7 & $2, 1$ & $0, 1.24751, 0.803639, 0.308613$ & $ [0.0491\pi, 1.9509\pi] $ & $ [0.9936\pi, 1.0064\pi] $ & $0.9445\pi$\\
    & 7 & $2, 1$ & $0, 0.752487, 1.19636, 1.69139$ & & & \\
D7b & 7 & $1, 2$ & $0.247513, 0.803639, 0.308613$ & $ [0.0260\pi, 1.9740\pi] $ & $ [0.9674\pi, 1.0326\pi] $ & $0.9414\pi$\\
    & 7 & $1, 2$ & $0, 0.752487, 0.308613, 0.803639$ & & & \\
\\
D9a & 9 & $3, 1$ & $0, 1.1113, 0.6905, 0.7503, 1.4422, 0.5945,$ & $ [0.1205\pi, 1.8795\pi] $ & $ [0.9901\pi, 1.0099\pi] $ & $0.8696\pi$\\
    & & & $1.8509, 1.6036, 0.0760$ & & &\\
D9b & 9 & $1, 3$ & $0, 0.8886, 0.4679, 0.4081, 1.1000, 1.9477,$ & $ [0.0227\pi, 1.9773\pi] $ & $ [0.9342\pi, 1.0658\pi] $ & $0.9116\pi$\\
    & & & $1.2041, 1.4515, 1.9238$ & & &\\
D11d\footnote{Possible D11a, D11b and D11c, respectively, for $(n_s, n_r)=(4,1)$, $(1,4)$ and $(3,2)$, derived by our method, are alternating pulses with wiggles, not relevant to rotation gates.} & 11 & $2, 3$ & $0, 1.9174, 0.9827, 1.2348, 0.7041, 0.1794,$ & $ [0.0307\pi, 1.9693\pi] $ & $ [0.9329\pi, 1.0671\pi] $ & $0.9021\pi$\\
    & & & $0.6527, 1.5332, 1.0226, 1.6844, 0.1870$ & & &\\
\\
\hline
\hline
\end{tabular}
\caption{
Phases of diversis passuum passband pulses for Hadamard gate. \\
}
\label{Table:PB-H-DN}
\end{table*}

\end{document}